\documentclass[preprint,12pt,number]{elsarticle}

\usepackage{setspace}
\usepackage{amssymb} 
\usepackage{amsmath}
\usepackage{graphicx}
\usepackage{subcaption}
\usepackage{float}
\usepackage{booktabs}
\usepackage{multirow}

\usepackage[a4paper, total={6in, 9in}]{geometry}

\journal{International Journal of Heat and Mass Transfer}

\begin{document}

\begin{frontmatter}

\title{}


\title{Analytical Modeling of Modulated Heating for Three-Dimensional Anisotropic Optical Thermometry of Multilayered Cylinders}

\author[label1]{William Spirnock\corref{cor1}} 
\author[label1]{Dihui Wang} 
\author[label1]{Heng Ban} 

\cortext[cor1]{Corresponding Author Email: wis57@pitt.edu}

\affiliation[label1]{organization={Department of Mechanical Engineering and Materials Science, University of Pittsburgh},
            addressline={3700 O'Hara St}, 
            city={Pittsburgh},
            postcode={15213}, 
            state={PA},
            country={USA}}

\begin{abstract}
Thermally relevant cylindrical geometries are used in a plethora of applications with thermal property differences depending significantly on material complexity. Manufacturing processes also induce substantial anisotropy ratios resulting in directionally dependent thermal transport. However, for cylindrical specimens, current thermal metrology resolves only isotropic thermal properties, assumes perfect thermal interfaces, and offers no formulation for temperature spatial dependence. In this work, we present an analytical framework that models the frequency dependent thermal transport of orthotropic arbitrary layered cylinders to be used as a forward model for existing optical thermometry techniques. This formulation captures the effects of non-negligible thermal boundary conductance and offers the first closed-form solution of a projected Gaussian laser, a common form of heating in thermal metrology, having a finite elliptic spot size on a curved surface. The novel anisotropic cylindrical model was compared with literature derivations involving two-dimensional isotropic transport for cylinders by applying simplifying assumptions. For single- and two-layer isotropic samples, the three-dimensional model agreed with established solutions with negligible error, while predictions for orthotropic multilayer samples agreed with finite-element simulations within 1.5\% across a range of material properties and laser geometries. Comparison with anisotropic planar models identified conditions under which a planar approximation can reduce computation time. These results establish the first forward model for noncontact characterization of anisotropic thermal properties in arbitrarily layered cylinders.

\end{abstract}


\end{frontmatter}

\section{Introduction}


Cylindrical and tubular geometries are ubiquitous across a broad range of scientific and engineering applications. In many cases, these systems exhibit inherently anisotropic thermal properties either as a result of their material structure or the manufacturing process. For example, cylindrical Li-ion cells are demonstrated to have significant anisotropies in thermal conductivities and thermal boundary conductance between the radial and axial directions due to the spiral assembly \cite{drake_measurement_2014,murashko_determination_2020}. Fiber-reinforced ceramic-matrix composite tubes such as silicon carbide (SiC) nuclear fuel cladding also demonstrate high anisotropy resulting from wound fiber orientation and thermal resistance between fibers \cite{deck_characterization_2015,deck_demonstration_nodate,wang_anisotropic_2022}. Thermal anisotropy is also prevalent in the development of additively manufactured (AM) components \cite{strumza_effect_2019,sarap_electrical_2024,shemelya_anisotropy_2017,malakizadi_post-processing_2022} where post-processing methods has a non-negligible impact on both material and thermal properties. Directionally dependent thermal transport is also observed in the case of molded and pyrolytic graphite cylinders, geological cores, and other orthotropic materials such as wood due to their inherently anisotropic structures \cite{popov_characterization_1999,vardi_thermal_1963,vay_anisotropic_2026}. Since each of these materials and structures are applied in fields of thermal relevance, quantifying their thermal properties, and how they vary directionally, is crucial for predicting heat flow, ensuring safe and controlled use, and guiding the design and fabrication of future cylindrical components.


Although cylindrical materials are widespread across various industries, current thermal characterization techniques are not mature for anisotropic measurements of non-planar geometries. Existing methods which consider cylindrical heating assume material isotropy and almost entirely one-dimensional heat flow with limited two-dimensional implementations. These methods are divided into two categories: contact and non-contact techniques. Many contact techniques design their experiment to ensure one-dimensional heat transfer. For instance, Fave et al designed a radial heat flow apparatus to determine the thermal conductivity of cylindrical samples in the radial direction \cite{fave_radial_2017}. Murashko et al utilized a similar principle in which a uniform heating layer was applied to the outside of a Li-ion cylindrical system and temperature measurements at specific regions allowed for the radial thermal conductivity to be determined \cite{murashko_determination_2020}. Contact methods have also been developed to determine the anisotropic thermal properties, as in the case of He et al and Drake et al, in which the heater/sensor configuration was designed to transfer heat in various directions depending on the configuration used \cite{he_method_2023,drake_measurement_2014}. Although reasonable accuracy was obtained, these methods require extensive manufacturing, are limited in the anisotropic properties that can be determined, and are prone to experimental errors arising from contact resistances.


Non-contact methods are often the preferred measurement technique as they are non-destructive, reduce common experimental errors arising from contact, and have the ability to perform spatially resolved measurements.
Such methods have been utilized to determine both isotropic and anisotropic thermal properties of planar specimens \cite{jiang_new_2022,wang_spatially_2024,wang_three-dimensional_2025,sheikh_photothermal_2025}, however, extensions to cylindrical geometries are limited. While the laser flash method is a commonly used non-contact technique, it is most often used to characterize the cross-plane thermal conductivity, especially in thin walled structures, and lacks the ability to probe at various depths to acquire information of each layer within a multilayer stack \cite{salazar_thermal_2024,duquesne_flash_2016,salazar_thermal_2011}. While certain variations of the flash method have demonstrated potential for anisotropic measurements of cylinders \cite{wang_anisotropic_2022}, the geometry was restricted to thin wall specimens, heat loss was assumed negligible, and FEM was used as opposed to a dedicated forward model which restricts the methods applicability to anisotropic complexity.

The most direct way to characterize the thermal properties of multilayered samples is to couple non-contact modulated optical characterization techniques like photothermal radiometry (PTR) with a forward model that incorporates distinct finite cylindrical layers. The use of frequency in methods like PTR enable selective depth probing to isolate the thermal effects of individual layers. Such methods have been developed and validated but only consider one-dimensional thermal transport. Wang et al developed a PTR methodology to characterize the radial thermal diffusivity of solid cylindrical samples \cite{wang_photothermal_2004}, and later extended it to a composite bilayer cylinder with a solid center \cite{wang_thermal-wave_2005}. Salazar et al generalized this framework for an arbitrary layered cylindrical shell with dissimilar properties using the thermal quadrupole method \cite{salazar_propagation_2006}. Xie et al utilized the Green's function method to derive a solution for temperature response of a multilayered cylinder due to a Gaussian laser spot \cite{xie_thermal_2011}. However, This model assumed a Gaussian intensity in the angular direction and a constant axial intensity rather than establishing a finite spot size. Xie's model also only derived the model in two dimensions ($r$ and $\phi$), assumed perfect contact between interfaces, isotropic layers, and did not consider spatial variation of the temperature distribution at various locations along the curvature. Although spatial variation has been considered, such as in Liu et al's work using lock-in thermography \cite{liu_quantitative_2020}, it was done using a uniform heating function, considered a single layer sample, and assumed isotropic thermal properties. There is a clear need for a model that incorporates each of these components to account for the complexity in novel cylindrical samples.


Various cylindrical geometry derivations exist in the literature that account for different levels of complexity. Some of these include multiple layers, multi-directional heat transport, and consideration of anisotropic transport phenomena. However, the majority make simplifying assumptions based on the intended application. In both Yang et al and Li et al, theoretical temperature solutions were derived for a multilayer cylindrical geometry \cite{yang_closed-form_2017,li_analytical_2013}. Both simplified to one-dimensional radial thermal transport and while Yang included thermal boundary resistance in their model, neither work considers anisotropy. Many theoretical models exist which employ two-directional heating, including both isotropic and anisotropic models \cite{rezaee_complete_2015,kayhani_general_2012,sarkar_analytical_2014, milosevic_analytical_2004,li_analytical_2013,yang_closed-form_2017,amiri_delouei_exact_2012,cossali_periodic_2009,lu_transient_2006}, however, there is often no treatment of the axial transport and many of these are solved using numerical methods as opposed to a closed form solution. Additionally, these solutions are commonly solved in the time-domain which reduces potential for depth-control measurements; those that do employ periodic heating are not in the form of a Gaussian laser spot. While certain models consider three-dimensional transport, they lack a comprehensive application of other qualities that are useful to consider in frequency-based thermal metrology \cite{amiri_delouei_comprehensive_2021,dalir_analytical_2014,abu-hijleh_analytical_1997,ozisik_transient_1980}. Although there are current models which incorporate different attributes of complexity, there exists no generalized solution that can be used for anisotropic multilayer cylindrical media that can be easily extended as a forward model in IR thermometry characterization techniques.


This work addresses these limitations. We apply the thermal quadrupole method to derive an analytical solution for a three-dimensional, non-axisymmetric forward model of orthotropic ($k_r \ne k_\phi \ne k_z$) multilayered cylinders for application in optical thermometry characterization techniques. For an arbitrary layered cylinder, this model allows for a distinct thermal conductivity tensor for an individual layer, thermal resistance across an interface, and a finite Gaussian heating spot in the angular and axial directions. The modulated Gaussian heating provides depth-profiling capabilities to assess the thermal response for a specific layer and results in distinct variations in the temperature distribution from the localized heating and detection. This novel theoretical model also implements spatially resolved detection through a built in offset in the Gaussian heating function. This will provide heightened characterization capabilities with regard to multi-parameter fitting of in-plane components as has been demonstrated in planar implementations \cite{feser_pump-probe_2014,tang_anisotropic_2021,rodin_simultaneous_2017,jiang_new_2022,wang_spatially_2024,wang_three-dimensional_2025}. Using simplifying assumptions, the forward model is first validated against the Xie et al angular Gaussian model \cite{xie_thermal_2011} for an isotropic single solid layer and an isotropic bilayer having a solid innermost layer. The cylindrical model was then validated against an existing planar model for a region that is comfortably within a planar approximation. The model was then validated across a range of FEM simulations encompassing various magnitudes of thermal conductivity, laser spot size, cylinder diameter, and thermal diffusion length. The cylindrical theory and planar model were then compared across various property combinations to provided guidance for identification of a planar approximation. These results establish a numerically validated forward model for future non-contact studies of complex cylindrical samples.


\section{Theory}

\subsection{Geometry, governing equation, and source}

\begin{figure}[H]
\centering

\begin{subfigure}[t]{0.49\textwidth}
\centering
\includegraphics[width=\linewidth]{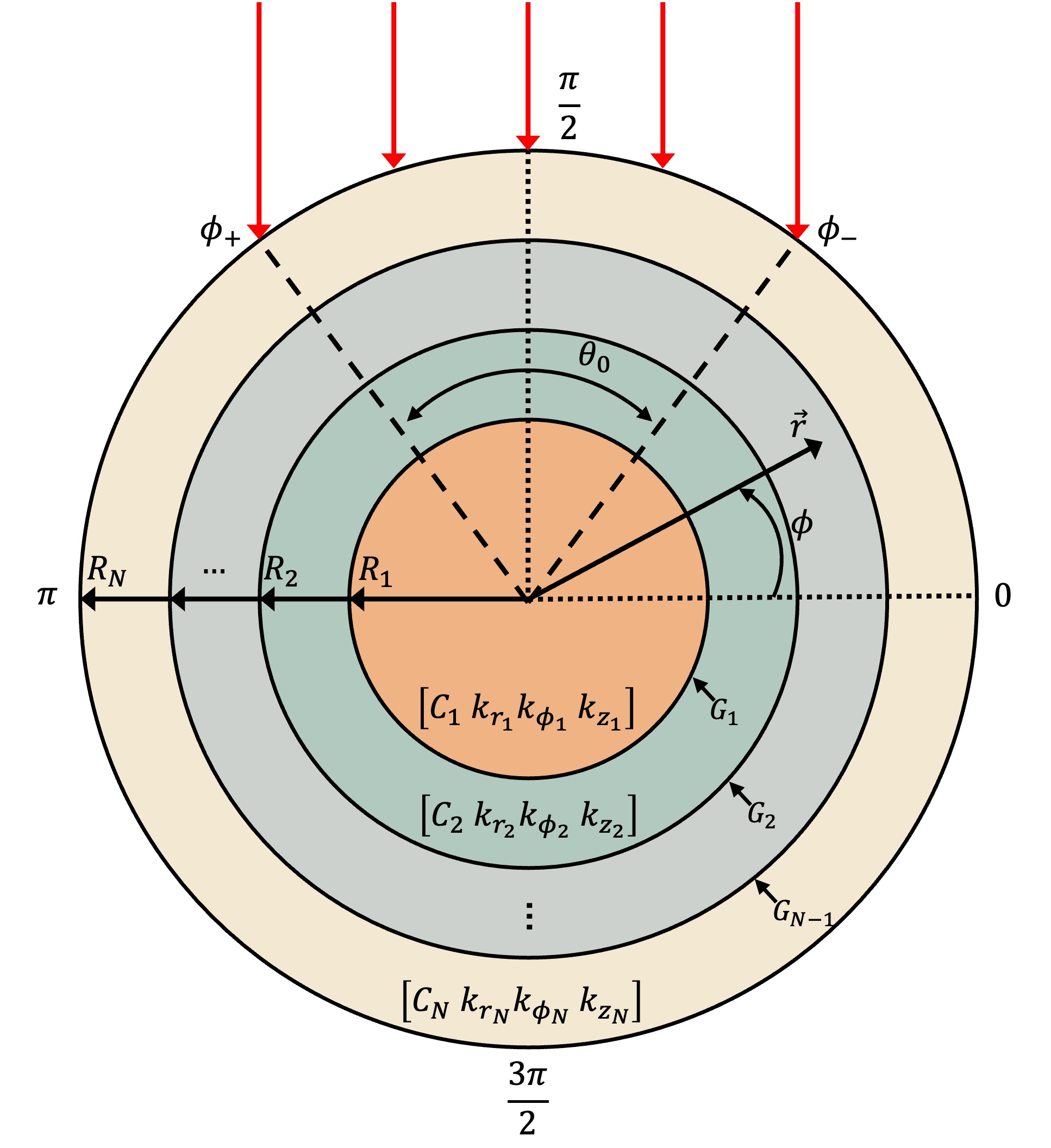}
\subcaption{}\label{fig:diagram_a}
\end{subfigure}
\begin{subfigure}[t]{0.49\textwidth}
\centering
\includegraphics[width=\linewidth]{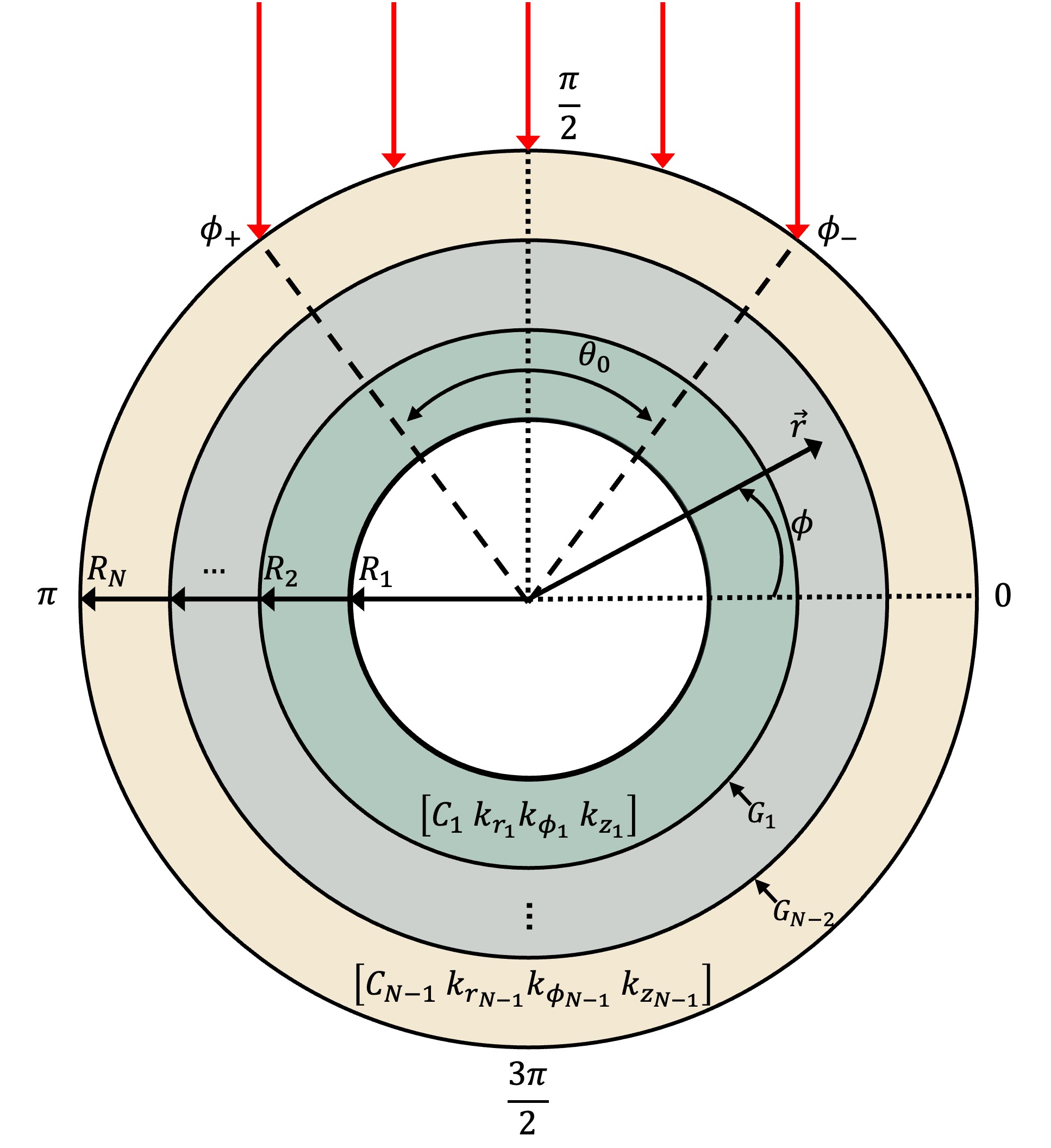}
\subcaption{}\label{fig:diagram_b}
\end{subfigure}\hfill
\caption{Schematic for laser-based heating of an arbitrary layered cylindrical geometry. The schematic applies to each of the three cases. (a) Case 1 would only apply to the solid core, $0 \leq r \leq R_1$. Case 2 applies to the full schematic, $0 \leq r \leq R_N$. (b) Case 3 applies to an arbitrary layer cylindrical shell, $R_1 \leq r \leq R_N$.}
\label{fig:schematic}
\end{figure}

Figure~\ref{fig:schematic} shows the cross-sectional model geometry for a solid cylinder and a cylindrical shell in the $r-\phi$ plane with positive axial ($z$) direction pointing out of the page. The azimuthal angle $\phi$ is measured counterclockwise from the $+x$ axis such that the top of the cylinder, the reference point for heating and detection, lies at $\phi = \frac{\pi}{2}$. Both geometries consist of an arbitrary number of concentric layers bounded by the radii $R_1 < R_2 < \dots < R_N$, with outer surface at $r = R_N$. Each layer $n$ has distinct thermal properties $[C \,\, k_r \,\, k_{\phi} \,\, k_z]_{n=1} \dots [C \,\, k_r \,\, k_{\phi} \,\, k_z]_{n=N}$ with directionally dependent thermal conductivities in the radial, circumferential, and axial directions. The interface between layers $n$ and $n + 1$ has a thermal boundary conductance $G_n$. The solid cylinder (Figure~\ref{fig:diagram_a}) has $N$ layers, the innermost being a solid core of radius $R_1$, and therefore $N - 1$ interfaces ($G_1, \dots , G_{N - 1}$). The cylindrical shell (Figure~\ref{fig:diagram_b} is air in the region $r < R_1$, meaning the wall contains $N - 1$ layers and $N - 2$ interfaces ($G_1, \dots, G_{N - 2}$).

We consider three cases: a solid cylinder, a multilayer solid cylinder with a solid core, and a multilayer cylindrical shell. Existing literature typically reports only a single case; the current work extends the forward model to both solid and hollow cylinders. Detailed derivations are provided in the supplementary materials document.

In each layer of the multilayered cylinder, $T_n(r,\phi,z,\omega)$ denotes the complex amplitude of the harmonic temperature rise relative to the ambient temperature, $T_\infty$, such that $T_{n}^{\mathrm{abs}}(r,\phi,z,t)=T_\infty+\operatorname{Re}\!\left[T_n(r,\phi,z,\omega)e^{i\omega t}\right]$, and satisfies the orthotropic heat-diffusion equation in cylindrical coordinates, where $n$ identifies the layer.

\begin{equation}
k_{r_n}
\left(
\frac{\partial^2 T_n}{\partial r^2}
+
\frac{1}{r}
\frac{\partial T_n}{\partial r}
\right)
+
k_{\phi_n}
\frac{1}{r^2}
\frac{\partial^2 T_n}{\partial \phi^2}
+
k_{z_n}
\frac{\partial^2 T_n}{\partial z^2}
=
C_n
\frac{\partial T_n}{\partial t}.
\label{eq:gov PDE}
\end{equation}

For each case, the outer surface ($r = R_N$) is heated by a modulated elliptical Gaussian laser beam that propagates perpendicular to the tangent at the top of the cylinder at $\phi = \frac{\pi}{2}$. The cartesian form of the heat flux due to an elliptic Gaussian laser can be mapped onto a cylindrical system by substituting $x=R_Ncos(\phi)$ and multiplying by a projection factor $sin(\phi)$ which accounts for the reduction in laser power per unit surface area caused by spreading over a large area at oblique incidence \cite{xie_thermal_2011,salazar_propagation_2006}. The heating function, $Q_H(R_N,\phi,z,t)$, is thus defined by both an angular and axial component

\begin{equation}
    Q_H(R_N,\phi,z,t)
=
\frac{P_0}{\Gamma}
\sin(\phi)
\exp\left[
-\frac{2\left(R_N\cos(\phi)-x_s\right)^2}{w_{x}^2}
\right]
\exp\left[
-\frac{2\left(z-z_s\right)^2}{w_z^2}
\right]
e^{i\omega t}
\label{eq:Q_H_t}
\end{equation}

\noindent{where $P_0$ is the absorbed harmonic amplitude, $w_x$ and $w_z$ are the elliptic $1/e^2$ radii, $x_s$ and $z_s$ are lateral offsets in the x-direction and z-direction, respectively, from the projected plane, and $\omega=2 \pi f$ is the angular frequency. The normalization constant, $\Gamma$, ensures that the integral of $Q_H$ on the illuminated surface is equal to the total absorbed power $\iint Q_H(R_N,\phi,z,t)\,R_N\,d\phi\,dz = P_0 e^{i\omega t}$.}

A spectral reduction is performed to transform Equation~\eqref{eq:gov PDE} into a solvable ODE. A temporal Fourier transform is applied to Equation~\eqref{eq:gov PDE} and Equation~\eqref{eq:Q_H_t} to shift from time domain to frequency domain using the convention ${F}^*(\omega)=\int_{-\infty}^{\infty}f(t)e^{-i\omega t} \,dt$. A complex Fourier series expansion in the azimuthal coordinate $\phi$ is utilized, $F(\phi) =\sum_{m=-\infty}^{\infty} \widetilde{F}(m)e^{im\phi}$, along with a Fourier transform for the $z-$direction, $\hat F(k)=\int_{-\infty}^{\infty}F(z)e^{-2i\pi kz} \,dz$ such that $T_n(r,\phi,z,t)) \rightarrow \hat{{\widetilde{T}}}^*(r,k,m,\omega)$. This reduces any given layer of the system to an ODE in $r$ where the fully transformed temperature is represented by $\hat{{\widetilde{T}}}^*(r)$,

\begin{equation}
r^2 \frac{d^2 \hat{{\widetilde{T}}}^*_n(r)}{dr^2}
+
r \frac{d \hat{{\widetilde{T}}}^*_n(r)}{dr}
-
\left(\beta_n^2 r^2 + \nu_n^2\right)
\hat{{\widetilde{T}}}^*_n(r)
\label{eq:ODE}
\end{equation}

where,

\begin{equation}
\beta_n^2(k,\omega) = \frac{k_{z_n}(2 \pi k)^2+i\omega C_n}{k_{r_n}}, \,\nu_n^2(m) = \frac{k_{\phi_n}}{k_{r_n}} |m|^2
\label{eq:beta_n}
\end{equation}

Equation~\eqref{eq:ODE} is a Bessel type ODE for a given layer with modified Bessel order $\nu_n = [{\frac{k_{\phi_n}}{k_{r_n}}|m|^2}]^{\frac{1}{2}}$ and has the solution

\begin{equation}
\hat{{\widetilde{T}}}^*_n(r) = F_n I_{\nu_n}(\beta_n r) + G_n K_{\nu_n}(\beta_n r)
\label{eq:Bessel ODE}
\end{equation}

where $I_{\nu_n}$ and $K_{\nu_n}$ are modified bessel functions of modified order $\nu_n$ for a given layer and $F_n$ and $G_n$ are layer-dependent coefficients which must be solved for. The transformed heating function becomes

\begin{equation}
\hat{\widetilde{Q}}^*_H(R_N)
=
\frac{P_0 \delta(\omega-\omega_0)}{\Gamma}
e^{-\frac{2(z-z_s)^2}{w_z^2}}
J_m \, ,
\qquad
\forall m
\label{eq:heat func transformed}
\end{equation}

\noindent{where ($\delta(\omega-\omega_0)$) represents the harmonic heating at a single angular frequency $\omega_0$. In Equation~\eqref{eq:heat func transformed}, the angular Fourier coefficient}

\begin{equation}
J_m = \int_{0}^{\pi} \sin(\phi) e^{-\frac{2\left(R_N\cos(\phi) - x_s\right)^2}{w_x^2}} e^{-im\phi}\, d\phi
\label{eq:angular Fourier coefficient}
\end{equation}

\noindent{quantifies the excitation strength of each circumferential mode $m$ by the projected Gaussian heat source. The integration is bounded from $0 \leq \phi \leq \pi$, such that the illumination cannot exceed a half cylinder. The decay of $J_m$ in the range $-M \leq m \leq M$ provides the criterion for selecting the azimuthal truncation order where the cylinder geometry as well as the angular width and shape of the heating profile influence the number of retained modes and the computational cost.}

\textit{Case 1: Solid cylinder}

A monolayer solid cylinder ($N=1$), with anisotropic thermal properties, is first considered.

\begin{equation}
\hat{{\widetilde{T}}}^*_1(r) = F_1 I_{\nu_1}(\beta_1 r) + G_1 K_{\nu_1}(\beta_1 r)
\label{eq:Bessel ODE solid}
\end{equation}

For a non-axisymmetric cylinder, the temperature at the centerline ($r=0$) must be finite and single-valued. Therefore, we must consider the temperature solution at all angular modes, $m$, using the regularity condition

\begin{equation}
\text{for } r=0,\qquad
\begin{cases}
\dfrac{\partial \hat{{\widetilde{T}}}^*_1(0)}{\partial r}=0,
& m=0, \\[8pt]
\hat{{\widetilde{T}}}^*_1(0)=0,
& m\neq 0.
\end{cases}
\label{eq:regularity diff}
\end{equation}

\begin{equation}
\hat{{\widetilde{T}}}^*_1(0)=\text{finite}
\label{eq:regularity temp}
\end{equation}

The surface of the cylinder ($r=R_1$) is subject to modulated heating from a Gaussian laser source and heat losses due to convection and radiation.

\begin{equation}
\hat{{\widetilde{Q}}}^*_1(R_1) = \hat{{\widetilde{Q}}}^*_H(R_1) - U\hat{{\widetilde{T}}}^*_1(R_1)
\label{eq:surface BC solid}
\end{equation}

\noindent{where the effective heat transfer coefficient combining the effects of convection and radiation is $U = h_c + 4 \epsilon \sigma T_\infty^3$. Here, $h_c$ is the convection heat transfer coefficient, $\epsilon$ is the surface emissivity, $T_\infty$ is the environmental temperature in Kelvin, and $\sigma = 5.67*10^{-8} \, W/m^2 K^4$ is the Stefan-Boltzmann constant. These boundary conditions are applied to Equation~\eqref{eq:Bessel ODE solid} to obtain the transformed temperature}

\begin{equation}
\hat{{\widetilde{T}}}^*_1(R_1) = \frac{I_{\nu_1}(\beta_1 R_1)}{k_{r_1} I_{\nu_1}(\beta_1 R_1) + U I_{\nu_1}(\beta_1 R_1)} \hat{{\widetilde{Q}}}^*_H(R_1)
\label{eq:transformed temp}
\end{equation}

\noindent{and upon applying the necessary inverse Fourier transforms for $\phi$ and $z$, the harmonic temperature solution in cylindrical coordinates is}

\begin{equation}
\begin{aligned}
T_1(R_1,\phi, z,\omega)
&=
\frac{
2 \sqrt{2}\,P_0\delta(\omega-\omega_0)
}{
w_x
\left[
\operatorname{erf}\!\left(\sqrt{\frac{2}{w_x^2}}(R_1-x_s)\right)
+
\operatorname{erf}\!\left(\sqrt{\frac{2}{w_x^2}}(R_1+x_s)\right)
\right]
}
\\
&\quad \times
\int_{-\infty}^{\infty}
\left(
\sum_{m=-\infty}^{\infty}
\hat{{\widetilde{T}}}^*_1(R_1)
J_m e^{im\phi}
\right)
e^{-\frac{\pi^2 k^2 w_z^2}{2}}
e^{2i\pi k(z-z_s)}
\, dk
\end{aligned}
\label{eq:temp final solid}
\end{equation}

\noindent{which gives the pointwise surface temperature of an anisotropic monolayer solid cylinder heated by an elliptical Gaussian source. Finite detection-area effects for optical thermometry are addressed in a later section.}


\textit{Case 2: Multilayer cylinder with solid core}

The layers surrounding the core are coupled using the thermal quadrupole method \cite{salazar_propagation_2006}, with the transformed radial heat flux defined as $\hat{{\widetilde{Q}}}_n^*=k_{r_n}\,d\hat{{\widetilde{T}}}_n^*/dr$ which is taken as a positive magnitude to account for the inward heat flux (negative $r$-direction). Within layer $n$, where $R_{n-1}\leq r\leq R_n$ and $n=2,\ldots,N$, the general solution of Equation~\eqref{eq:Bessel ODE} relates the temperature and heat flux at the inner and outer surfaces through the layer matrix $\mathbf{M}_n$:

\begin{equation}
\begin{bmatrix}
\hat{{\widetilde{T}}}_n^*(R_n) \\
\hat{{\widetilde{Q}}}_n^*(R_n)
\end{bmatrix}
=
\mathbf{M}_n
\begin{bmatrix}
\hat{{\widetilde{T}}}_n^*(R_{n-1}) \\
\hat{{\widetilde{Q}}}_n^*(R_{n-1})
\end{bmatrix},
\qquad
\mathbf{M}_n
=
\begin{bmatrix}
A_n & B_n \\
C_n & D_n
\end{bmatrix},
\label{eq:layer matrix}
\end{equation}

\noindent{whose coefficients depend on $\nu_n$, $\beta_n$, and the layer radii and are provided in the supplementary materials document. At each interface, $r=R_n$, the heat flux is continuous and the temperature decreases by $\hat{{\widetilde{Q}}}_n^*(R_n)/G_n$ from the outer side to the inner side, such that}

\begin{equation}
\begin{bmatrix}
\hat{{\widetilde{T}}}_{n+1}^*(R_n) \\
\hat{{\widetilde{Q}}}_{n+1}^*(R_n)
\end{bmatrix}
=
\boldsymbol{\mathcal{G}}_n
\begin{bmatrix}
\hat{{\widetilde{T}}}_n^*(R_n) \\
\hat{{\widetilde{Q}}}_n^*(R_n)
\end{bmatrix},
\qquad
\boldsymbol{\mathcal{G}}_n
=
\begin{bmatrix}
1 & 1/G_n \\
0 & 1
\end{bmatrix}.
\label{eq:interface matrix}
\end{equation}

The multilayer case is coupled to the solid cylinder by utilizing the Case 1 solution as a boundary condition for the innermost surface of the second layer at $R_2$. Starting from the core surface and proceeding outward by alternating the interface and layer matrices gives

\begin{equation}
\begin{aligned}
\begin{bmatrix}
\hat{{\widetilde{T}}}^*(R_N) \\
\hat{{\widetilde{Q}}}^*(R_N)
\end{bmatrix}
&=
\mathbf{M}_N
\boldsymbol{\mathcal{G}}_{N-1}
\mathbf{M}_{N-1}
\boldsymbol{\mathcal{G}}_{N-2}
\cdots
\boldsymbol{\mathcal{G}}_2
\mathbf{M}_2
\boldsymbol{\mathcal{G}}_1
\begin{bmatrix}
\hat{{\widetilde{T}}}_1^*(R_1) \\
\hat{{\widetilde{Q}}}_1^*(R_1)
\end{bmatrix}
\\[6pt]
&\equiv
\begin{bmatrix}
A & B \\
C & D
\end{bmatrix}
\begin{bmatrix}
\hat{{\widetilde{T}}}_1^*(R_1) \\
\hat{{\widetilde{Q}}}_1^*(R_1)
\end{bmatrix},
\end{aligned}
\label{eq:overall transfer matrix}
\end{equation}

\noindent{where the non-subscript quantities denote the outer-surface values and $A$, $B$, $C$, and $D$ are the elements of the overall transfer matrix which contain information regarding the geometry and thermal properties of the various layers. Note the matrix $\boldsymbol{\mathcal{G}}_{1}$} must be to the right of the second layer matrix $\mathbf{M}_2$, to correctly represent the imperfect thermal interface between the solid cylinder (serving as a boundary condition) and first added layer.

Applying the solid cylinder surface conditions along with a mixed boundary condition for Gaussian heating and convective/radiative losses at the outermost layer, $\hat{{\widetilde{Q}}}^*(R_N) = \hat{{\widetilde{Q}}}^*_H(R_N) - U\hat{{\widetilde{T}}}^*(R_N)$, the temperature solution at $r=R_N$ is

\begin{equation}
\begin{aligned}
\hat{{\widetilde{T}}}^*(R_N)
=
\left[
1+
\frac{
\left(
A I_{\nu_1}(\beta_1 R_1)
+
B k_{r_1} I'_{\nu_1}(\beta_1 R_1)
\right)U
}{
C I_{\nu_1}(\beta_1 R_1)
+
D k_{r_1} I'_{\nu_1}(\beta_1 R_1)
}
\right]^{-1}
\\
\times
\frac{
A I_{\nu_1}(\beta_1 R_1)
+
B k_{r_1} I'_{\nu_1}(\beta_1 R_1)
}{
C I_{\nu_1}(\beta_1 R_1)
+
D k_{r_1} I'_{\nu_1}(\beta_1 R_1)
}
\hat{{\widetilde{Q}}}^*_H(R_N)
\end{aligned}
\label{eq:transformed temp solid + arb layer}
\end{equation}

\noindent{and applying inverse transforms to Equation~\eqref{eq:transformed temp solid + arb layer} yields a similar result to Equation~\eqref{eq:temp final solid}}

\begin{equation}
\begin{aligned}
T(R_N,\phi,z,\omega)
&=
\frac{
2 \sqrt{2}\,P_0\delta(\omega-\omega_0)
}{
w_x
\left[
\operatorname{erf}\!\left(\sqrt{\frac{2}{w_x^2}}(R_N-x_s)\right)
+
\operatorname{erf}\!\left(\sqrt{\frac{2}{w_x^2}}(R_N+x_s)\right)
\right]
}
\\
&\quad \times
\int_{-\infty}^{\infty}
\left(
\sum_{m=-\infty}^{\infty}
\hat{{\widetilde{T}}}^*(R_N)
J_m e^{im\phi}
\right)
e^{-\frac{\pi^2 k^2 w_z^2}{2}}
e^{2i\pi k(z-z_s)}
\, dk
\end{aligned}
\label{eq:temp final solid + arb}
\end{equation}

\noindent{which represents the pointwise temperature response at the surface of an arbitrary layered anisotropic cylinder which has a solid cylinder core.}


\textit{Case 3: Multilayer cylindrical shell}

An arbitrary layered cylindrical shell is solved in a manner similar to Case 2. The thermal quadrupole method is used to construct a transfer matrix relating the temperature and heat flux at the inner and outer surfaces. Case 3 differs from Case 2 only in its inner boundary condition. Using the indexing shown in Figure~\ref{fig:diagram_b}, layer $n$ occupies $R_n \leq r \leq R_{n+1}$. Using the layer and interface matrices defined in Equations~\eqref{eq:layer matrix} and~\eqref{eq:interface matrix}, respectively, the shell transfer matrix takes the form of Equation~\eqref{eq:overall transfer matrix}, with the matrix product $\mathbf{M}_{N-1}\boldsymbol{\mathcal{G}}_{N-2}\cdots\boldsymbol{\mathcal{G}}_1\mathbf{M}_1$. The product begins with $\mathbf{M}_1$ because no interface exists at $R_1$ for a cylindrical shell. This formulation can be solved by applying Equation~\eqref{eq:surface BC solid} at $r = R_N$ and a convective/radiative heat loss boundary condition ar $r = R_1$

\begin{equation}
\begin{aligned}
\hat{{\widetilde{Q}}}^*(R_1) = U\hat{{\widetilde{T}}}^*(R_1)
\end{aligned}
\label{eq:R1 BC}
\end{equation}

resulting in the transformed temperature solution for an arbitrary layered cylindrical shell

\begin{equation}
\begin{aligned}
\hat{{\widetilde{T}}}^*(R_N) = \frac{A + UB}{C + U(A + D + UB)}
\hat{{\widetilde{Q}}}^*_H(R_N)
\end{aligned}
\label{eq:transformed temp shell + arb layer}
\end{equation}

\noindent{The inverse transformed temperature solution defined by Equation~\eqref{eq:temp final solid + arb} applies to Case 2 by substituting Equation~\eqref{eq:transformed temp shell + arb layer} instead of Equation~\eqref{eq:transformed temp solid + arb layer}. A detailed derivation for each case including definitions of all coefficients are provided in the supplementary materials document.}

\subsection{Detection Window}

The derived temperature solutions for each of the cases are for a point on the cylinder surface. However, in actuality, optical thermal characterization methods like PTR \cite{sheikh_photothermal_2025}, frequency domain thermoreflectance (FDTR) \cite{tang_anisotropic_2021}, and time-domain thermoreflectance (TDTR) \cite{jiang_tutorial_2018}, compute an average temperature using either a finite sized or Gaussian detection window. In some cases, the heating source is designed to be several times larger than the detection window \cite{wang_three-dimensional_2025, wang_spatially_2024} to meet the criteria for point detection. But, in general, the dimensions of the detection must be accounted for. Two main categories of optical thermometry are IR-based techniques which utilize an IR detector or camera to capture a thermal response and thermoreflectance techniques which require a continuous wave Gaussian laser for measurement of the temperature change. In thermoreflectance, the measured temperature response is predicated on a specular surface such that the probe laser can be reflected from the sample surface and directed into a photodetector for measurement purposes. However, in the case of a curved surface, this condition of specularity is unlikely to occur unless the cylinder diameter is significantly larger than the probe, in which case a planar approximation would apply. Therefore, this work only considers IR-based detection methods, which do not have the same condition for specularity.

\subsubsection{Detector-area averaging}

IR-detection techniques collect the emitted IR signal over a finite detection window, typically a rectangular or elliptical region. Assuming uniform detector sensitivity per unit projected area and sufficiently small temperature variations for the radiance--temperature relationship to be linearized, the measured harmonic response can be represented as a detector-weighted spatial average of the surface temperature. In the present model, the Gaussian heat source may be displaced laterally by $(x_s,z_s)$, while the detection window remains centered at $\phi=\pi/2$ and $z=0$. Although the detection window is fixed in the Cartesian projection plane, its corresponding footprint on the cylindrical surface is distorted by the surface curvature. Therefore, a geometric treatment similar to that used for the heating spot is required. The detector-weighted temperature is defined as the ratio of the sensitivity-weighted temperature response to the total detector sensitivity:

\begin{equation}
\overline{T}(R_N,\omega)
=
\displaystyle \iint T(R_N,z,\phi,\omega)\,W(z,\phi)\,dA
\label{eq:temp weighted general}
\end{equation}

\noindent{where the differential surface area element for a cylindrical geometry is $dA = R_N d\phi dz$, and the normalized weight function is defined as}

\begin{equation}
W(z,\phi) = \frac{W_0(z,\phi)}{N_{IR}}
\label{eq:norm weighting function}
\end{equation}

\noindent{Here we assume a Lambertian emitter, whose radiance is independent of the viewing angle. The power that a surface element sends to the detector then scales with its projected area as seen by the detector, so the weight is $W_0(z,\phi)=sin(\phi)$ inside of the detection window and $W_0(z,\phi)=0$ outside. The normalization constant $N_{IR} = \iint W_0(z,\phi)\,R_N d\phi dz$ is to ensure the normalized weighting function $W(\phi, z)$ integrates to unity over the detection area, $\iint W(z,\phi)\,R_N d\phi dz = 1$, such that the resulting temperature represents a normalized detector-weighted temperature response rather than a quantity that depends on the absolute magnitude of the weighting function.}

For an elliptical detection window, the integration bounds are determined by mapping the window from the Cartesian projection plane onto the cylindrical surface. The detection window is centered at $x=0$ and $z=0$, corresponding to $\phi=\pi/2$ on the cylindrical surface. Substituting $x=R_N\cos(\phi)$ into the boundary of the projected ellipse, $\frac{x^2}{w_{x,\mathrm{IR}}^2}+\frac{z^2}{w_{z,\mathrm{IR}}^2}=1,$ produces two angular bounds at each axial position $z$. These bounds correspond to the lateral edges of the elliptical detection window and are denoted by $\phi_{\mathrm{IR},-}(z)$ and $\phi_{\mathrm{IR},+}(z)$:

\begin{equation}
\phi_{\mathrm{IR},\pm}(z)
=
\cos^{-1}
\mp
\frac{w_{x,\mathrm{IR}}}{R_N}
\sqrt{
1-\frac{z^2}{w_{z,\mathrm{IR}}^2}
}.
\label{eq:IR-angular-bounds}
\end{equation}

Thus, the IR weighted average temperature response at the surface of the cylinder for a circular detection area is

\begin{equation}
\overline{T}(R_N,\omega)
=
\frac{1}{\pi w_{x_{IR}}w_{z_{IR}}}
\int_{-w_{z_{IR}}}^{w_{z_{IR}}}
\int_{\phi_{IR_-}(z)}^{\phi_{IR_+}(z)}
T(R_N,\phi,z,\omega)\,
\sin(\phi)\,
R_N\,d\phi\,dz
\label{eq:IR_averaged_temperature circular}
\end{equation}

\noindent{where $w_{x_{IR}}$ and $w_{z_{IR}}$ are the finite detector radii in the $x$ and $z$ direction from the projection plane. And for a rectangular detection region, these bounds are no longer dependent on $z$}

\begin{equation}
\begin{aligned}
\overline{T}(R_N,\omega)
=
\frac{1}{4 w_{x_{IR} w_{z_{IR}}}}
\int_{-w_{z_{IR}}}^{w_{z_{IR}}}
\int_{\phi_{IR_-}}^{\phi_{IR_+}}
T(R_N,\phi,z,\omega)\,
\sin(\phi)\,
R_N\,d\phi\,dz
\label{eq:IR_averaged_temperature square}
\end{aligned}
\end{equation}

\section{Numerical validation}

Validation of the present theory is performed for various cases. The solid cylinder single and bilayer derivation are compared to an isotropic system with dissimilar layers from a published axisymmetric derivation \cite{xie_thermal_2011}. The proposed model requires simplifying assumptions to be compatible with the Xie et al model. The phase and normalized amplitude are compared between the models with negligible error. The cylindrical shell model derived in this work is also compared to a multilayer anisotropic planar model with appropriate parameters selected such that the cylindrical theory is within the planar approximation. The phase and normalized amplitude again show negligible error between the models. These comparisons serve as initial benchmarks for the full cylindrical theory under various assumptions prior to validation of more complex cases. Although amplitude and phase are both compared, the phase is of greater interest as it is generally considered a more reliable means of thermal property estimation as opposed to amplitude which can require additional calibration \cite{corona_recent_2023}. Note that for all validation cases, the convective and radiative losses were set to zero.

\subsection{Isotropic case}

The present theory is compared with a model derived explicitly for isotropic thermal property characterization of solid cylindrical specimens \cite{xie_thermal_2011}. Xie et al defines their model in a $(r,\phi)$ coordinate system in which there is Gaussian heating only in the x-direction which is projected onto a cylindrical surface, as done in this work. They assume a uniform flux in the axial direction of the cylinder which, in our model, can be represented as a Gaussian spot with significant ellipticity $\frac{w_z}{w_x} \rightarrow \infty$. It was determined that there is negligible change in the temperature solution when $\frac{w_z}{w_x} \geq 1000$ and so 1000 was used in this study. A spot size of $w_x = 1 \, mm$ was used to simulate various cases in which the thermal conductivity and ratio between the spot size and cylinder radius $(w_x / R_N)$ were altered. This ratio ranges from zero to one and is a primary factor in assessing the curvature effects and whether or not a planar model can be applied. A large ratio means curvature effects dominate, but cannot exceed one since the angular expansion integration is defined on the interval $[0,\pi]$. The cylinder radius was calculated from three chosen ratios of 0.3, 0.6, 0.9 to represent solutions from small to significant curvature effects. A volumetric heat capacity of $C = 1*10^6 \, Jm^{-3}K^{-1}$ was selected for all samples.

\begin{figure}[H]
\centering

\begin{subfigure}[t]{0.49\textwidth}
\centering
\includegraphics[width=\linewidth]{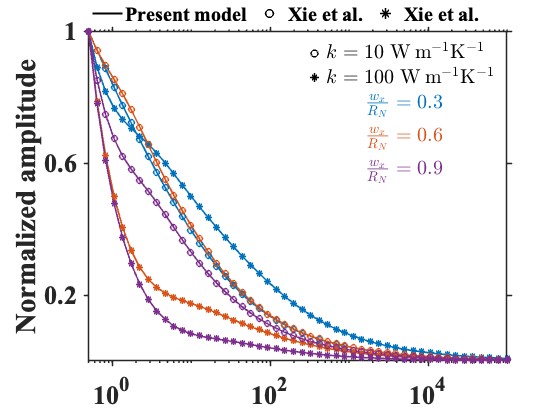}
\subcaption{}\label{fig:IV_top_left}
\end{subfigure}
\begin{subfigure}[t]{0.49\textwidth}
\centering
\includegraphics[width=\linewidth]{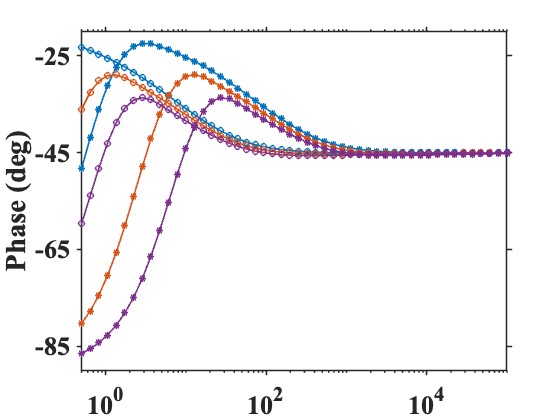}
\subcaption{}\label{fig:IV_top_right}
\end{subfigure}\hfill
\begin{subfigure}[t]{0.49\textwidth}
\centering
\includegraphics[width=\linewidth]{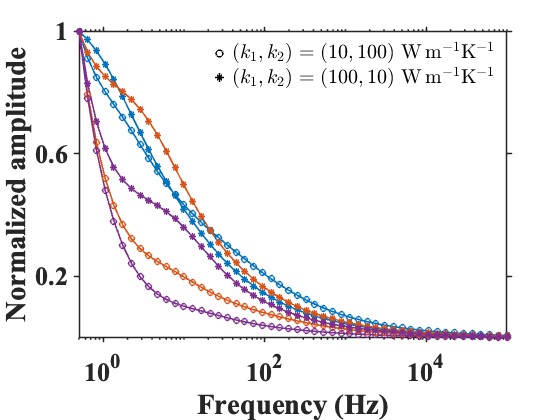}
\subcaption{}\label{fig:IV_bottom_left}
\end{subfigure}
\begin{subfigure}[t]{0.49\textwidth}
\centering
\includegraphics[width=\linewidth]{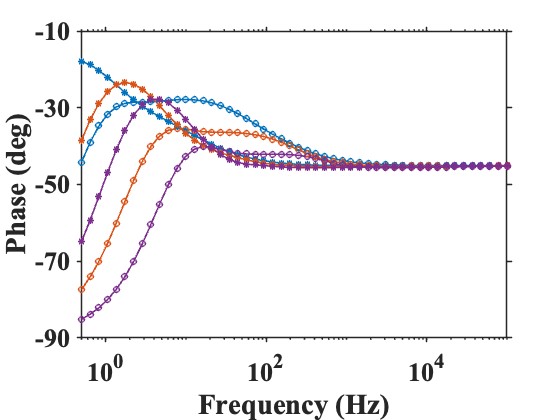}
\subcaption{}\label{fig:IV_bottom_right}
\end{subfigure}
\caption{Verification against isotropic analytical model of Xie et al \cite{xie_thermal_2011}. (a) Normalized amplitude and (b) phase for a monolayer isotropic solid cylinder. (c) Normalized amplitude and (d) phase for a bilayer isotropic solid cylinder with dissimilar layers.}
\label{fig:Isotropic Validation}
\end{figure}

Case 1 and Case 2 were compared to Xie et al's bilayer model. For the solid cylinder case, the material was assumed to be the same for each layer in \cite{xie_thermal_2011}. In the bilayer comparison, the thermal boundary conductance in the present model was set to an exceedingly high value $(G \rightarrow \infty)$ in which effects from interfacial resistance are negligible. Figures~\ref{fig:IV_top_left} and~\ref{fig:IV_top_right} evaluate the normalized amplitude and phase for six conditions formed from two thermal conductivities $k = 10 \, Wm^{-1}K^{-1}$ and $k = 100 \, Wm^{-1}K^{-1}$. Each thermal conductivity tests the three ratios of Gaussian spot to cylinder radius. These values are selected to span a range of thermal properties and curvature effects. Figures~\ref{fig:IV_bottom_left} and~\ref{fig:IV_bottom_right} perform the same analysis for two conductivity pairs $(k_1, k_2) = (10,100) \, Wm^{-1}K^{-1}$ and $(100, 10) \, Wm^{-1}K^{-1}$ where $k_1$ and $k_2$ are the thermal conductivities of the solid core and outer shell layer, respectively. Note that the thicknesses of the outer layer in the bilayer case is calculated such that the two layers are equal, i.e. $R_1 = R_2 - \frac{R_2}{2}$, where $R_1$ is the radius of the solid cylinder and $R_2$ is the outer radius of the cylindrical shell layer.

A negligible error, $(<0.01 \, \%)$, was obtained when comparing the present model to the Xie et al solution when the angular mode number satisfied the truncation requirements. From the phase plots, it is evident that as $\frac{w_x}{R_N}$ increases, the phase trend shifts to higher frequencies. The cylinder outer radius is calculated based on the Gaussian radius and selected ratio of spot size to cylinder radius. Therefore a larger ratio would result in a smaller cylinder radius and thus greater penetration of the thermal diffusion length at higher frequencies relative to the material thickness. The conductivity also changes the phase trends consistently with prior observations from planar model results. A larger thermal conductivity results in a longer thermal diffusion length leading to thermal property information over a larger range of frequencies.


\subsection{Anisotropic case}

\subsubsection{Planar approximation}


The anisotropic cylindrical model was compared to a planar model of similar complexity for further validation. The isotropic validation case was based on the application of simplifying assumptions to the present theory to make the model compatible with previous work. The same principle can be applied to compare an existing planar model to the new cylindrical theory within a regime of planar approximation. Various ratios of spot size and cylinder radii were tested to identify a region that the present theory satisfies a planar approximation across a range of thermal conductivities. It was found that $\frac{w}{R_N} = 0.02$ displayed little variation between the planar and cylindrical model regardless of thermal conductivity. Any ratio smaller than this value demanded extreme truncation requirements and increased computational run time substantially.

\begin{figure}[H]
\centering

\begin{subfigure}[t]{0.49\textwidth}
\centering
\includegraphics[width=\linewidth]{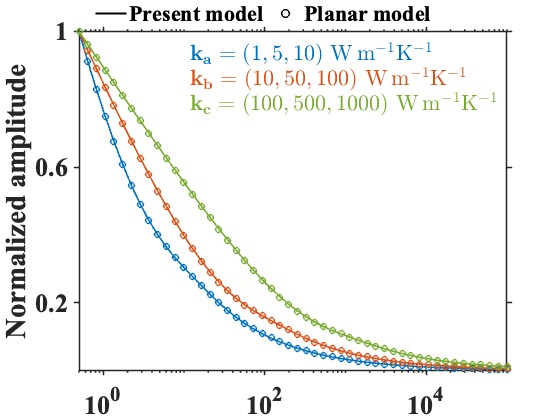}
\subcaption{}\label{fig:amp single}
\end{subfigure}
\begin{subfigure}[t]{0.49\textwidth}
\centering
\includegraphics[width=\linewidth]{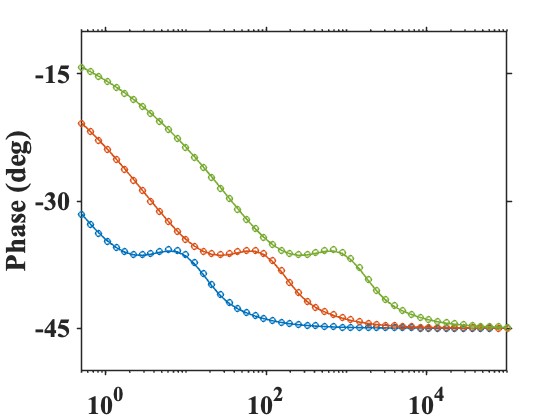}
\subcaption{}\label{fig:phase single}
\end{subfigure}
\begin{subfigure}[t]{0.49\textwidth}
\centering
\includegraphics[width=\linewidth]{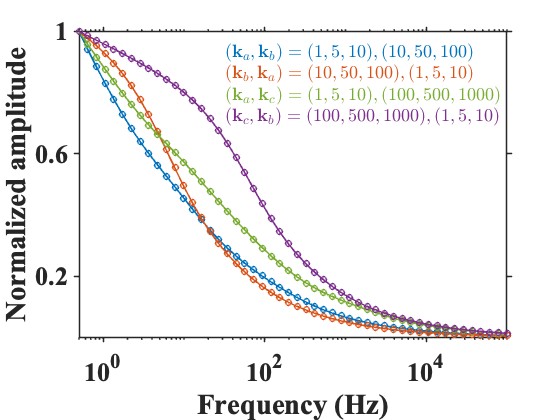}
\subcaption{}\label{fig:amp double}
\end{subfigure}
\begin{subfigure}[t]{0.49\textwidth}
\centering
\includegraphics[width=\linewidth]{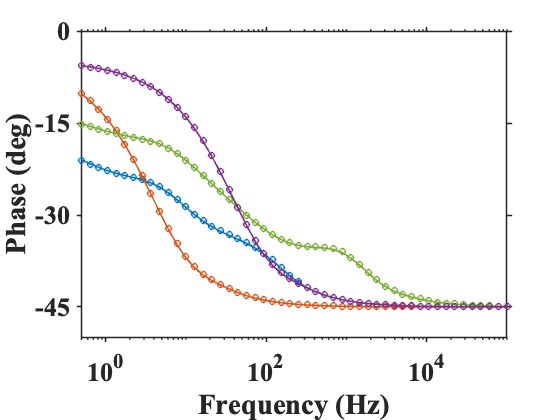}
\subcaption{}\label{fig:phase double}
\end{subfigure}
\caption{Verification against anisotropic planar model within a region of planar approximation, $\frac{w}{R_N}$. (a) Normalized amplitude and (b) phase for an anisotropic monolayer. (c) Normalized amplitude and (d) phase for an anisotropic bilayer with dissimilar layers.}
\label{fig:Anisotropic Planar Validation}
\end{figure}

Figures~\ref{fig:amp single} and~\ref{fig:phase single}, show the resulting normalized amplitude and phase for a single cylindrical shell for three different cases of anisotropy spanning a large range of thermal conductivity. When comparing the cylindrical and planar coordinate system, we use $k_{r} = k_z$, $k_{\phi} = k_x$, and $k_z = k_y$. Note that in the cylindrical coordinate system, $z$ refers to the axial direction along the cylinder length, but in the cartesian coordinate system it represents the depth direction through the material. Excellent agreement is obtained between the planar and cylindrical models with overall error within $0.4 \, \%$. The error trend is approximately the same for each property combination with larger thermal properties corresponding to a shift in error toward higher frequencies for similar reasons as in the isotropic validation case. Figures~\ref{fig:amp double} and~\ref{fig:phase double} show the normalized amplitude and phase for a bilayer cylindrical shell for various combinations of dissimilar layers. The models match reasonably well with slightly higher error compared to the single layer case. The larger error associated with the planar comparison is due to the selected approximation region. For an ideal approximation, as $\frac{w}{R_N} \rightarrow 0$ it follows that $E_\phi (\%) \rightarrow 0$. However, as $\frac{w}{R_N}$ decreases, computational complexity increases substantially. The agreement in the planar limit validates the cylindrical theory within an important region and demonstrates that the complexity of the cylindrical model simplifies to the planar solution. This will be expanded upon in a later section.















\subsubsection{FEM results}

A three-dimensional finite element method (FEM) code was developed in python using the scikit-fem library to validate the analytical model for cases in which the planar approximation is invalid. The code solves the frequency-domain heat equation with independent radial, circumferential, and axial thermal conductivities. The domain was discretized using first-order hexahedral elements, with frequency-dependent mesh refinement based on the thermal diffusion length with reduced mesh sizes near the heating source and detection region. A thermal boundary conductance was applied at each interface ($G=50*10^6 \frac{W}{m^2 K}$). A Gaussian heat flux was applied to the outermost layer of the cylindrical surface where periodic conditions were enforced circumferentially, and all unheated regions were set to adiabatic boundary conditions.

\begin{figure}[H]
\centering

\begin{subfigure}[t]{0.49\textwidth}
\centering
\includegraphics[width=\linewidth]{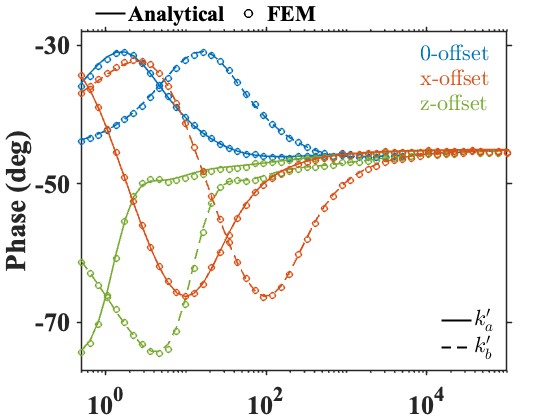}
\subcaption{}\label{fig:mono solid}
\end{subfigure}
\begin{subfigure}[t]{0.49\textwidth}
\centering
\includegraphics[width=\linewidth]{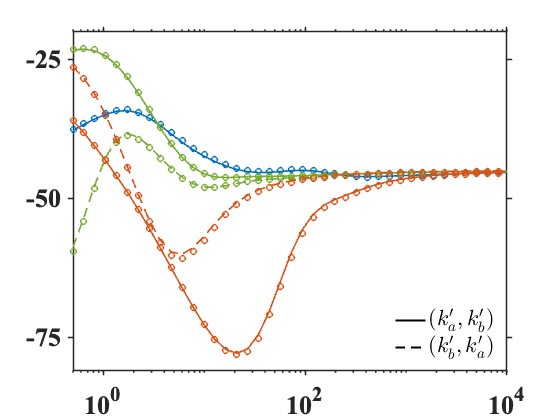}
\subcaption{}\label{fig:bilayer solid}
\end{subfigure}
\begin{subfigure}[t]{0.49\textwidth}
\centering
\includegraphics[width=\linewidth]{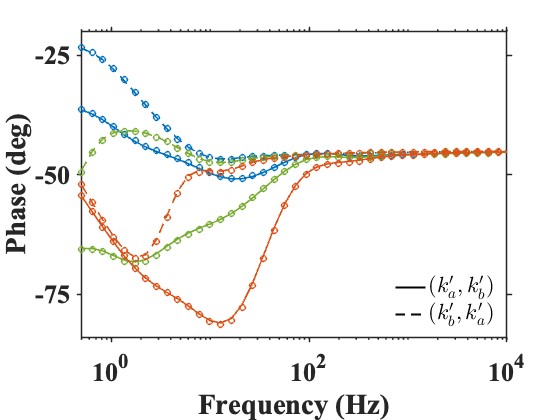}
\subcaption{}\label{fig:bilayer shell}
\end{subfigure}
\begin{subfigure}[t]{0.49\textwidth}
\centering
\includegraphics[width=\linewidth]{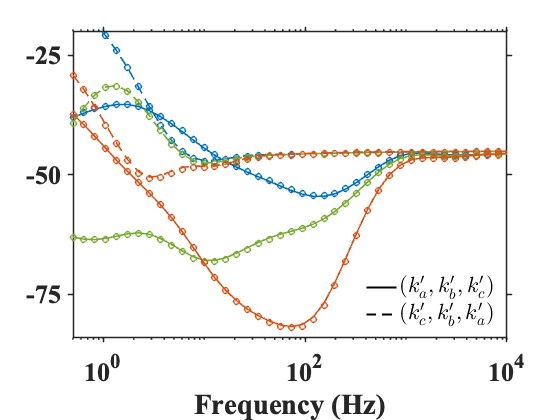}
\subcaption{}\label{fig:trilayer shell}
\end{subfigure}

\caption{Numerical verification against FEM for an anisotropic (a) monolayer solid cylinder, (b) bilayer solid cylinder, (c) bilayer cylindrical shell, and (d) trilayer cylindrical shell.}
\label{fig:Anisotropic Validation}
\end{figure}

This code was used to generate data for three offset cases (0-offset, x-offset, z-offset) and four different layer configurations. Each case assumes $w_x = w_z = w
$. These included a monolayer solid cylinder, a bilayer cylinder with a solid core, a bilayer cylindrical shell, and a trilayer cylindrical shell. The resulting FEM phase was compared with the analytical model for various combinations of dissimilar layers. Comparison of the FEM and analytical models demonstrated that increased effects of curvature occur when $k_r > k_{\phi} > k_z$. This is due to the increased significance of the geometry when the radial penetration depth is non-negligible relative to the cylinder radius, meaning when $\frac{\mu_r}{R_N}$ is appreciable. The same reasoning follows for the thermal diffusion length of the circumferential direction compared to the axial direction. Thus, the conductivity ordering promotes thermal transport in the radial and circumferential directions while restricting axial heat spreading, causing the temperature field to interact more strongly with the radius-dependent terms of the cylindrical heat equation. A ratio of $\frac{w}{R_N} = 0.9$ was also used for all cases to further ensure curvature effects dominate the thermal response. We utilize the same combinations of thermal conductivity as $k_a , k_b , k_c$ but flip the thermal conductivity values to ensure the ordering condition is satisfied, i.e. $k_a' = (10,5,1) \frac{W}{mK}, k_b' = (100,50,10) \frac{W}{mK}, k_c' = (1000,500,100) \frac{W}{mK}$. Additional information regarding parameter selection for Figure~\ref{fig:Anisotropic Validation} can be found in the supplementary materials document.

Reasonable agreement was found between the analytical model derived in this work and the FEM model with a maximum error of 1.51\% occurring at a specific frequency in the sweep for the single layer solid cylinder case. For the bilayer solid cylinder, the maximum error was 1.28\% and the bilayer and trilayer shells had a maximum error of 0.84\% and 0.85\%, respectively. This study numerically verifies the forward model for application to characterization of thermally relevant cylindrical geometries.


\section{Planar limit}

As discussed in previous sections, when certain conditions are satisfied, the cylindrical model can be approximated using a planar model. This approximation may be more desirable for thermal characterization of cylindrical specimens primarily to reduce computational cost for data fitting. It was found that the cylindrical solution code computes for a substantially longer amount of time; however, a nonlinear least squares curve fitting code may demand even more computational complexity.

\begin{figure}[H]
\centering

\begin{subfigure}[t]{0.32\textwidth}
\centering
\includegraphics[width=\linewidth]{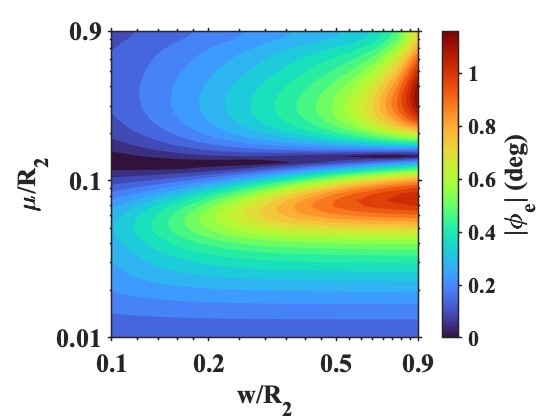}
\subcaption{}\label{fig:LR0.1}
\end{subfigure}
\begin{subfigure}[t]{0.32\textwidth}
\centering
\includegraphics[width=\linewidth]{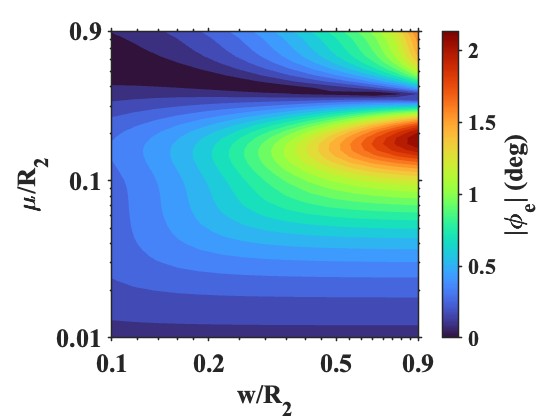}
\subcaption{}\label{fig:LR0.3}
\end{subfigure}
\begin{subfigure}[t]{0.32\textwidth}
\centering
\includegraphics[width=\linewidth]{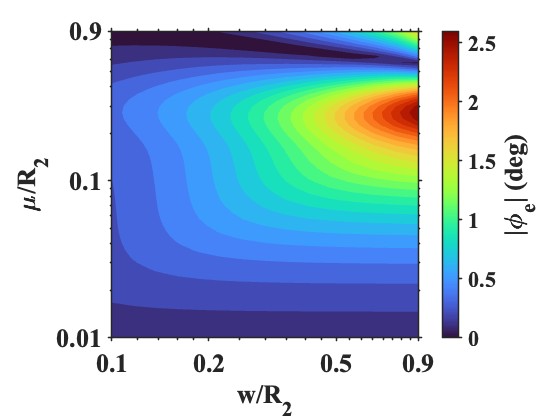}
\subcaption{}\label{fig:LR0.5}
\end{subfigure}
\begin{subfigure}[t]{0.32\textwidth}
\centering
\includegraphics[width=\linewidth]{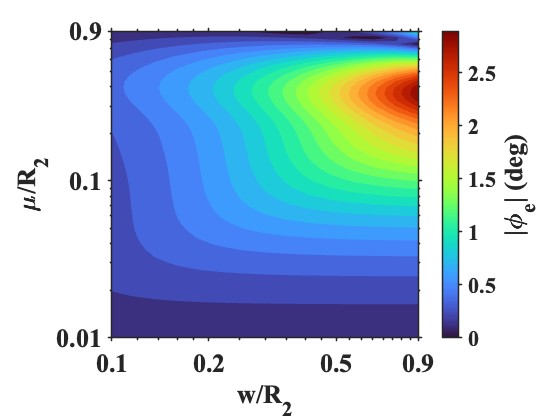}
\subcaption{}\label{fig:LR0.7}
\end{subfigure}
\begin{subfigure}[t]{0.32\textwidth}
\centering
\includegraphics[width=\linewidth]{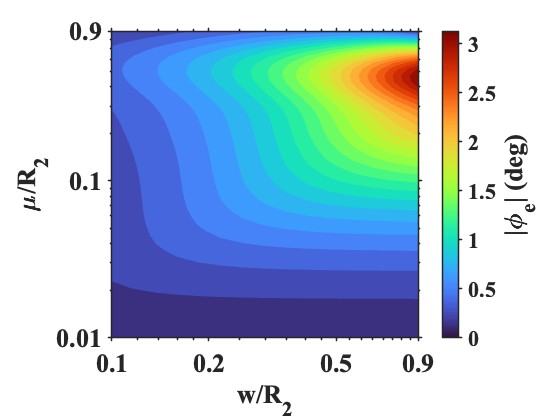}
\subcaption{}\label{fig:LR0.9}
\end{subfigure}

\caption{Contour plots illustrating the dependency of error between the cylindrical and planar model, $\phi_e$, on various dimensionless quantities such as $\frac{\mu}{R_2}$, $\frac{w}{R_2}$, across different magnitudes of $\frac{L}{R_2}$ being (a) $\frac{L}{R_2} = 0.1$, (b) $\frac{L}{R_2} = 0.3$, (c) $\frac{L}{R_2} = 0.5$, (d) $\frac{L}{R_2} = 0.7$, (e) $\frac{L}{R_2} = 0.9$.}
\label{fig:Planar Limit Contour}
\end{figure}

An isotropic monolayer cylindrical shell was analyzed to provide a rigorous foundation for identification of the planar limit in Figure~\ref{fig:Planar Limit Contour}. After extensive numerical analysis, which involved tests across various combinations of thermal properties and geometric parameters, it was found that the error between the cylindrical and planar models, $\phi_e = \phi_{cyl} - \phi_{planar}$, is dependent on three ratios involving the cylinder outer radius, $R_2$. These ratios are between the thermal diffusion length and outer radius ($\frac{\mu}{R_2}$), the laser spot size and outer radius ($\frac{w}{R_2}$), and the shell thickness and outer radius ($\frac{L}{R_2}$). These measure the effect of curvature for the primary length dependencies. Figures~\ref{fig:Planar Limit Contour}(a)--(e) show the variation of model error as $\frac{L}{R_2}$ increases. Across all ratios of $\frac{L}{R_2}$, the lowest error occurs when both $\frac{\mu}{R_2}$ and $\frac{w}{R_2}$ are minimized. As $\frac{L}{R_2}$ increases, the maximum error also increases in the red regions indicating that thicker cylinders will result in larger error between models. This implies that the solid cylinder with arbitrary layers will be more prone to curvature effects. More specifically, thicker cylinders, larger laser spot sizes, and lower frequency regions increase the error between the cylindrical and planar models where the thermal diffusion length is related to frequency via $\mu = \sqrt\frac{k}{\pi C f}$. Thus, by having geometric information for the laser and cylinder to be measured, as well as a predicted value for thermal conductivity and volumetric heat capacity, a frequency range can be selected for thermal characterization of a cylindrical geometry using the planar model, since in this range, ($k_r, k_{\phi}, k_z$) in the cylindrical system is equivalent to ($k_z, k_x, k_y$) in the planar system.

The established relationship for the model error of an isotropic monolayer shell, $\phi_e(\frac{w}{r_2},\frac{\mu}{r_2},\frac{L}{r_2})$, has only three dependencies. However, for anisotropic and multilayered specimens, the phase error becomes increasingly complex. For example, the phase error of an orthotropic bilayer shell would be $\phi_e(\frac{w}{r_3},\frac{\mu_{r_1}}{r_3},\frac{\mu_{\phi_1}}{r_3},\frac{\mu_{z_1}}{r_3},\frac{\mu_{r_2}}{r_3},\frac{\mu_{\phi_2}}{r_3},\frac{\mu_{z_2}}{r_3},\frac{L_1}{r_3},\frac{L_2}{r_3})$. This does not even take into consideration a beam offset, elliptical laser spot, or the influence of the thermal boundary conductance and how that compounds with additional layers. Therefore, it is unreasonable to establish this relationship for more complex cases, but it is still crucial to understand how these parameters impact $\phi_e$ to determine if the planar model is applicable.

\begin{figure}[H]
\centering

\begin{subfigure}[t]{0.32\textwidth}
\centering
\includegraphics[width=\linewidth]{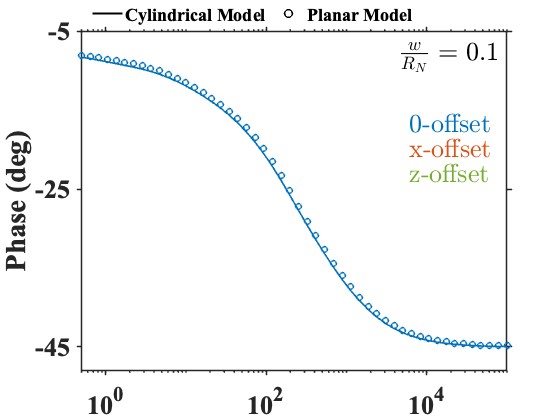}
\subcaption{}\label{fig:0offset small}
\end{subfigure}
\begin{subfigure}[t]{0.32\textwidth}
\centering
\includegraphics[width=\linewidth]{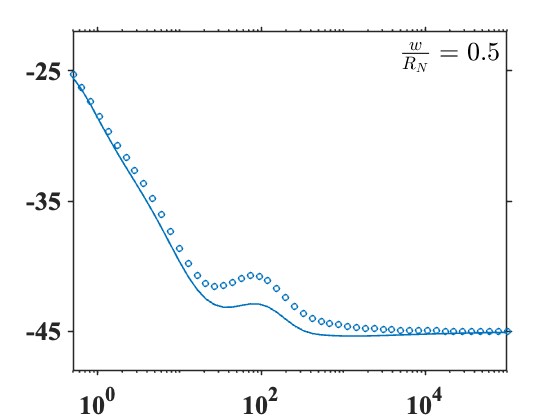}
\subcaption{}\label{fig:0offset med}
\end{subfigure}
\begin{subfigure}[t]{0.32\textwidth}
\centering
\includegraphics[width=\linewidth]{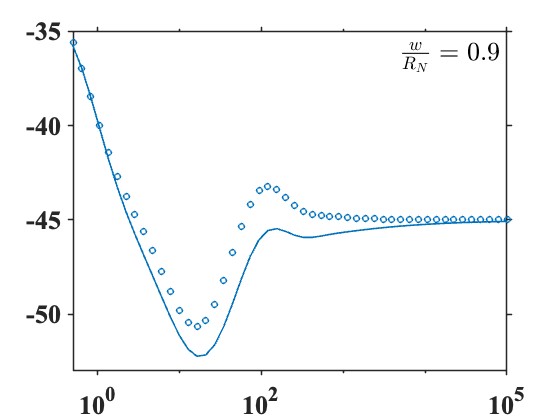}
\subcaption{}\label{fig:0offset large}
\end{subfigure}
\begin{subfigure}[t]{0.32\textwidth}
\centering
\includegraphics[width=\linewidth]{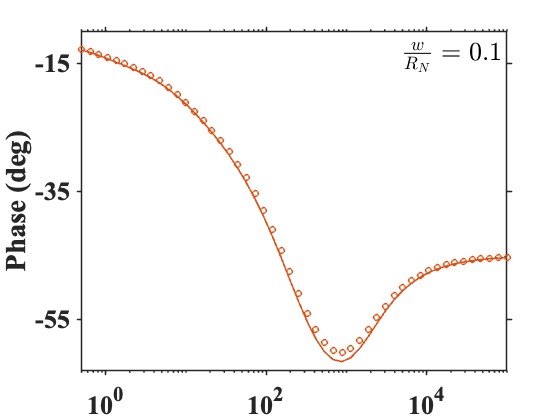}
\subcaption{}\label{fig:xoffset small}
\end{subfigure}
\begin{subfigure}[t]{0.32\textwidth}
\centering
\includegraphics[width=\linewidth]{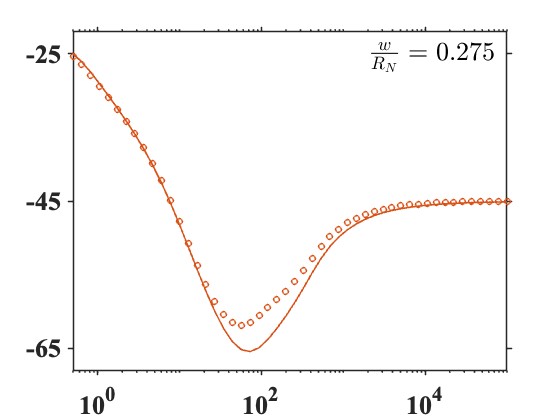}
\subcaption{}\label{fig:xoffset med}
\end{subfigure}
\begin{subfigure}[t]{0.32\textwidth}
\centering
\includegraphics[width=\linewidth]{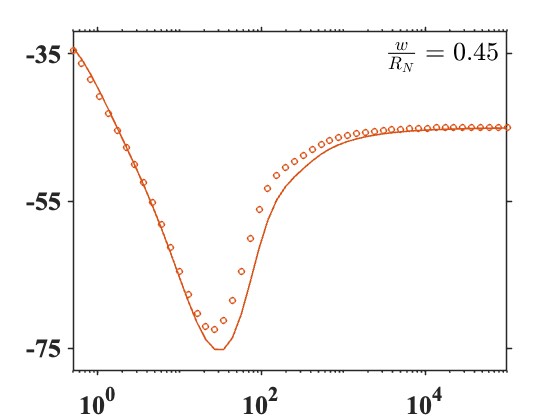}
\subcaption{}\label{fig:xoffset lare}
\end{subfigure}
\begin{subfigure}[t]{0.32\textwidth}
\centering
\includegraphics[width=\linewidth]{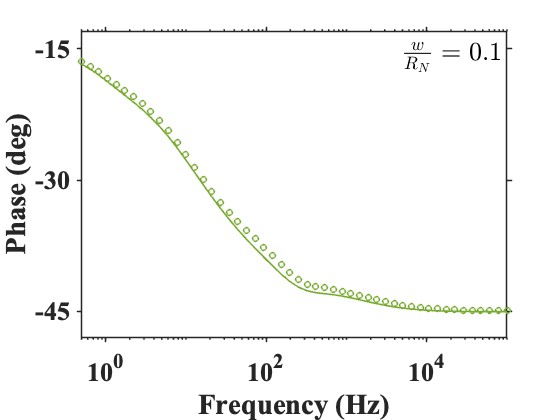}
\subcaption{}\label{fig:zoffset small}
\end{subfigure}
\begin{subfigure}[t]{0.32\textwidth}
\centering
\includegraphics[width=\linewidth]{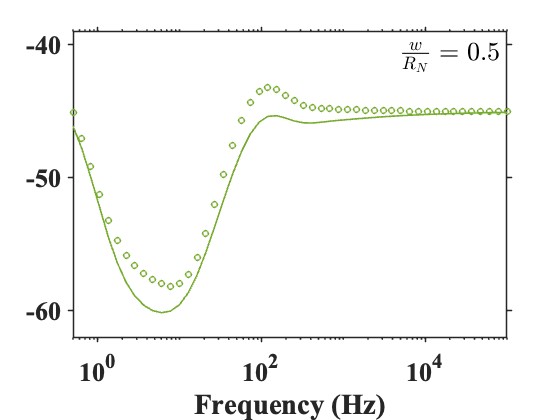}
\subcaption{}\label{fig:zoffset med}
\end{subfigure}
\begin{subfigure}[t]{0.32\textwidth}
\centering
\includegraphics[width=\linewidth]{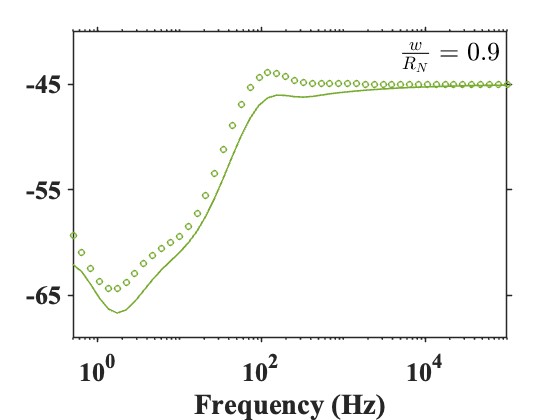}
\subcaption{}\label{fig:zoffset lare}
\end{subfigure}
\caption{Comparison between the anisotropic cylindrical and planar models for various combinations of $\frac{w}{R_N}$ and different offsets. (a) Zero-offset condition for $\frac{w}{R_N} = 0.1$, (b) $\frac{w}{R_N} = 0.5$, (c) $\frac{w}{R_N} = 0.9$. (d) X-offset condition for $\frac{w}{R_N} = 0.1$, (e) $\frac{w}{R_N} = 0.275$, (f) $\frac{w}{R_N} = 0.45$. (g) Z-offset condition for $\frac{w}{R_N} = 0.1$, (h) $\frac{w}{R_N} = 0.5$, (i) $\frac{w}{R_N} = 0.9$.}
\label{fig:Planar Limit Line}
\end{figure}

A beam offset is commonly applied in planar-based optical thermometry as a means to determine orthotropic thermal conductivities. Thus, it is crucial to determine if a planar model can still be used as an approximation for an anisotropic cylinder. A trilayer cylindrical shell with dissimilar anisotropic layers was tested for three cases: zero-offset, x-offset, z-offset. The thermal conductivity tensor for the layers from outermost layer to innermost are ($k_c',k_b',k_a'$). The overall thickness of the cylinder was based on the condition $\frac{L}{R_N} = 0.9$. These criteria were selected to base the analysis on a sample highly susceptible to the effect of curvature. For each test, the offset magnitude was set to be equal to the magnitude of the laser radius. From Figure~\ref{fig:Planar Limit Contour}, it is clear that $\frac{w}{R_N}$ can be selected for any $\frac{\mu}{R_N}$ and $\frac{L}{R_N}$ to obtain a planar approximation. An offset was performed over various magnitudes of $\frac{w}{R_N}$ to determine if the spot size can be selected in an extreme case that would allow for use of a planar model over a cylindrical model. Figures~\ref{fig:Planar Limit Line}(a)--(i) show that for all offset cases, regardless of layer number or anisotropy, a spot size can be selected to ensure the planar approximation holds. Therefore, the ratio of spot size to cylinder outer radius is the most significant constraint for utilizing the planar model. Regardless of the system in question, Figure~\ref{fig:Planar Limit Contour} can still serve as a useful guide to parameter selection if it is desired to use a planar model for property fitting over the cylindrical model derived in this work.

\section{Summary and outlook}

Cylindrical geometries are utilized in a variety of thermally relevant applications. These samples often have intrinsic or artificially induced anisotropic thermal properties which result in non-uniform heat spreading and so must be accurately characterized to ensure proper operation with regard to safety and efficiency. However, there is a significant limitation having to do with the characterization of these complex cylindrical geometries. To bridge this gap, this work developed a first of its kind multi-case derivation for a three-dimensional anisotropic forward model of an arbitrary layered cylinder for use in the field of optical thermometry.

The impact of this work is multifaceted. The derivation provides a direct means to determine the applicability of a planar solution for measuring the anisotropic thermal properties of cylindrical specimens. By comparing the two solutions for a particular geometry and set of properties, the appropriate laser spot size can be selected to perform thermal characterization using a planar approximation, thereby reducing the computational cost when fitting the experimental data. For complex structures such as fiber-based composite cylinders, this model can be used as a forward model in the case where an effective anisotropic thermal conductivity tensor is desired or where there are system-based limitations with regard to the diameter of the heating laser spot. This work also applied two of the most common detection methods (in various geometries) used in optical thermometry for the scenario in which there is a detection spot with non-negligible sensitivity.

The derived model was validated using existing solutions of reduced complexity and FEM simulations to verify the analytical model's applicability over a range of geometric parameters and thermal property combinations. However, this study does not establish parameter identifiability, uncertainty propagation, or analyze how percent deviations between the planar and cylindrical models equates to nonlinear least squares curve fitting error. These features will be analyzed in future work. Nonetheless, the established forward model provides the foundation for its implementation in inverse fitting models for anisotropic arbitrary layered cylindrical shells and solids.



\bibliographystyle{ieeetr}
\bibliography{References/Batteries,References/SiC, References/others,References/AM,References/contact_techniques,References/noncontact_techniques,References/IR_Techniques,References/Cylindrical_Theory}

@article{strumza_effect_2019,
	title = {The effect of texture on the anisotropy of thermophysical properties of additively manufactured {AlSi10Mg}},
	volume = {29},
	issn = {22148604},
	url = {https://linkinghub.elsevier.com/retrieve/pii/S2214860419301800},
	doi = {10.1016/j.addma.2019.06.013},
	language = {en},
	urldate = {2026-08-28},
	journal = {Additive Manufacturing},
	author = {Strumza, Einat and Yeheskel, Ori and Hayun, Shmuel},
	month = oct,
	year = {2019},
	pages = {100762},
}

@article{sarap_electrical_2024,
	title = {Electrical and {Thermal} {Anisotropy} in {Additively} {Manufactured} {AlSi10Mg} and {Fe}-{Si} {Samples}},
	volume = {13},
	issn = {2075-1702},
	url = {https://www.mdpi.com/2075-1702/13/1/1},
	doi = {10.3390/machines13010001},
	language = {en},
	number = {1},
	urldate = {2026-08-28},
	journal = {Machines},
	author = {Sarap, Martin and Tiismus, Hans and Kallaste, Ants and Saarna, Mart and Kolnes, Märt and Shams Ghahfarokhi, Payam and Vaimann, Toomas},
	month = dec,
	year = {2024},
	pages = {1},
}

@article{shemelya_anisotropy_2017,
	title = {Anisotropy of thermal conductivity in {3D} printed polymer matrix composites for space based cube satellites},
	volume = {16},
	issn = {22148604},
	url = {https://linkinghub.elsevier.com/retrieve/pii/S2214860417302324},
	doi = {10.1016/j.addma.2017.05.012},
	language = {en},
	urldate = {2026-08-28},
	journal = {Additive Manufacturing},
	author = {Shemelya, Corey and De La Rosa, Angel and Torrado, Angel R. and Yu, Kevin and Domanowski, Jennifer and Bonacuse, Peter J. and Martin, Richard E. and Juhasz, Michael and Hurwitz, Frances and Wicker, Ryan B. and Conner, Brett and MacDonald, Eric and Roberson, David A.},
	month = aug,
	year = {2017},
	pages = {186--196},
}

@article{malakizadi_post-processing_2022,
	title = {Post-processing of additively manufactured metallic alloys – {A} review},
	volume = {179},
	issn = {08906955},
	url = {https://linkinghub.elsevier.com/retrieve/pii/S0890695522000591},
	doi = {10.1016/j.ijmachtools.2022.103908},
	language = {en},
	urldate = {2026-08-28},
	journal = {International Journal of Machine Tools and Manufacture},
	author = {Malakizadi, Amir and Mallipeddi, Dinesh and Dadbakhsh, Sasan and M'Saoubi, Rachid and Krajnik, Peter},
	month = aug,
	year = {2022},
	pages = {103908},
}

@article{drake_measurement_2014,
	title = {Measurement of anisotropic thermophysical properties of cylindrical {Li}-ion cells},
	volume = {252},
	issn = {03787753},
	url = {https://linkinghub.elsevier.com/retrieve/pii/S0378775313019502},
	doi = {10.1016/j.jpowsour.2013.11.107},
	language = {en},
	urldate = {2026-08-27},
	journal = {Journal of Power Sources},
	author = {Drake, S.J. and Wetz, D.A. and Ostanek, J.K. and Miller, S.P. and Heinzel, J.M. and Jain, A.},
	month = apr,
	year = {2014},
	pages = {298--304},
}

@article{murashko_determination_2020,
	title = {Determination of the through-plane thermal conductivity and specific heat capacity of a {Li}-ion cylindrical cell},
	volume = {162},
	issn = {00179310},
	url = {https://linkinghub.elsevier.com/retrieve/pii/S001793102033266X},
	doi = {10.1016/j.ijheatmasstransfer.2020.120330},
	language = {en},
	urldate = {2026-08-27},
	journal = {International Journal of Heat and Mass Transfer},
	author = {Murashko, K.A. and Pyrhönen, J. and Jokiniemi, J.},
	month = dec,
	year = {2020},
	pages = {120330},
}

@article{li_analytical_2013,
	title = {Analytical solution to heat conduction in finite hollow composite cylinders with a general boundary condition},
	volume = {60},
	issn = {00179310},
	url = {https://linkinghub.elsevier.com/retrieve/pii/S0017931013000525},
	doi = {10.1016/j.ijheatmasstransfer.2013.01.029},
	language = {en},
	urldate = {2026-08-21},
	journal = {International Journal of Heat and Mass Transfer},
	author = {Li, Min and Lai, Alvin C.K.},
	month = may,
	year = {2013},
	pages = {549--556},
}

@article{yang_closed-form_2017,
	title = {Closed-form analytical solutions of transient heat conduction in hollow composite cylinders with any number of layers},
	volume = {108},
	issn = {00179310},
	url = {https://linkinghub.elsevier.com/retrieve/pii/S001793101632035X},
	doi = {10.1016/j.ijheatmasstransfer.2016.12.020},
	language = {en},
	urldate = {2026-08-21},
	journal = {International Journal of Heat and Mass Transfer},
	author = {Yang, Bingen and Liu, Shibing},
	month = may,
	year = {2017},
	pages = {907--917},
}

@article{rezaee_complete_2015,
	title = {Complete anisotropic time-dependent heat equation in {KTP} crystal under repetitively pulsed {Gaussian} beams: a numerical approach},
	volume = {54},
	issn = {1559-128X, 2155-3165},
	shorttitle = {Complete anisotropic time-dependent heat equation in {KTP} crystal under repetitively pulsed {Gaussian} beams},
	url = {https://opg.optica.org/abstract.cfm?URI=ao-54-6-1241},
	doi = {10.1364/AO.54.001241},
	language = {en},
	number = {6},
	urldate = {2026-08-02},
	journal = {Applied Optics},
	author = {Rezaee, Mostafa Mohammad and Sabaeian, Mohammad and Motazedian, Alireza and Jalil-Abadi, Fatemeh Sedaghat and Khaldi-Nasab, Ali},
	month = feb,
	year = {2015},
	pages = {1241},
}

@article{kayhani_general_2012,
	title = {A general analytical solution for heat conduction in cylindrical multilayer composite laminates},
	volume = {52},
	issn = {12900729},
	url = {https://linkinghub.elsevier.com/retrieve/pii/S1290072911002626},
	doi = {10.1016/j.ijthermalsci.2011.09.002},
	language = {en},
	urldate = {2026-08-21},
	journal = {International Journal of Thermal Sciences},
	author = {Kayhani, M.H. and Norouzi, M. and Amiri Delouei, A.},
	month = feb,
	year = {2012},
	pages = {73--82},
}

@article{sarkar_analytical_2014,
	title = {Analytical modeling of temperature distribution in an anisotropic cylinder with circumferentially-varying convective heat transfer},
	volume = {79},
	issn = {00179310},
	url = {https://linkinghub.elsevier.com/retrieve/pii/S0017931014007583},
	doi = {10.1016/j.ijheatmasstransfer.2014.08.060},
	language = {en},
	urldate = {2026-08-21},
	journal = {International Journal of Heat and Mass Transfer},
	author = {Sarkar, D. and Shah, K. and Haji-Sheikh, A. and Jain, A.},
	month = dec,
	year = {2014},
	pages = {1027--1033},
}

@article{milosevic_analytical_2004,
	title = {Analytical solution of transient heat conduction in a two-layer anisotropic cylindrical slab excited superficially by a short laser pulse},
	volume = {47},
	copyright = {https://www.elsevier.com/tdm/userlicense/1.0/},
	issn = {00179310},
	url = {https://linkinghub.elsevier.com/retrieve/pii/S0017931003006136},
	doi = {10.1016/j.ijheatmasstransfer.2003.10.023},
	language = {en},
	number = {8-9},
	urldate = {2026-08-02},
	journal = {International Journal of Heat and Mass Transfer},
	author = {Milošević, N.D. and Raynaud, M.},
	month = apr,
	year = {2004},
	pages = {1627--1641},
}

@article{amiri_delouei_exact_2012,
	title = {Exact analytical solution of unsteady axi-symmetric conductive heat transfer in cylindrical orthotropic composite laminates},
	volume = {55},
	copyright = {https://www.elsevier.com/tdm/userlicense/1.0/},
	issn = {00179310},
	url = {https://linkinghub.elsevier.com/retrieve/pii/S0017931012002554},
	doi = {10.1016/j.ijheatmasstransfer.2012.04.012},
	language = {en},
	number = {15-16},
	urldate = {2026-08-21},
	journal = {International Journal of Heat and Mass Transfer},
	author = {Amiri Delouei, A. and H. Kayhani, M. and Norouzi, M.},
	month = jul,
	year = {2012},
	pages = {4427--4436},
}

@article{cossali_periodic_2009,
	title = {Periodic heat conduction in a solid homogeneous finite cylinder},
	volume = {48},
	copyright = {https://www.elsevier.com/tdm/userlicense/1.0/},
	issn = {12900729},
	url = {https://linkinghub.elsevier.com/retrieve/pii/S129007290800121X},
	doi = {10.1016/j.ijthermalsci.2008.05.009},
	language = {en},
	number = {4},
	urldate = {2026-08-21},
	journal = {International Journal of Thermal Sciences},
	author = {Cossali, G.E.},
	month = apr,
	year = {2009},
	pages = {722--732},
}

@article{lu_transient_2006,
	title = {Transient analytical solution to heat conduction in multi-dimensional composite cylinder slab},
	volume = {49},
	copyright = {https://www.elsevier.com/tdm/userlicense/1.0/},
	issn = {00179310},
	url = {https://linkinghub.elsevier.com/retrieve/pii/S0017931005005703},
	doi = {10.1016/j.ijheatmasstransfer.2005.08.033},
	language = {en},
	number = {5-6},
	urldate = {2026-08-21},
	journal = {International Journal of Heat and Mass Transfer},
	author = {Lu, X. and Tervola, P. and Viljanen, M.},
	month = mar,
	year = {2006},
	pages = {1107--1114},
}

@article{amiri_delouei_comprehensive_2021,
	title = {A {Comprehensive} {Review} on {Multi}-{Dimensional} {Heat} {Conduction} of {Multi}-{Layer} and {Composite} {Structures}: {Analytical} {Solutions}},
	volume = {30},
	issn = {1003-2169, 1993-033X},
	shorttitle = {A {Comprehensive} {Review} on {Multi}-{Dimensional} {Heat} {Conduction} of {Multi}-{Layer} and {Composite} {Structures}},
	url = {https://link.springer.com/10.1007/s11630-021-1517-1},
	doi = {10.1007/s11630-021-1517-1},
	language = {en},
	number = {6},
	urldate = {2026-08-21},
	journal = {Journal of Thermal Science},
	author = {Amiri Delouei, Amin and Emamian, Amin and Sajjadi, Hasan and Atashafrooz, Meysam and Li, Yueming and Wang, Lian-Ping and Jing, Dengwei and Xie, Gongnan},
	month = nov,
	year = {2021},
	pages = {1875--1907},
}

@article{dalir_analytical_2014,
	title = {Analytical {Solution} of the {Problem} on the {Three}-{Dimensional} {Transient} {Heat} {Conduction} in a {Multilayer} {Cylinder}},
	volume = {87},
	copyright = {http://www.springer.com/tdm},
	issn = {1062-0125, 1573-871X},
	url = {http://link.springer.com/10.1007/s10891-014-0988-2},
	doi = {10.1007/s10891-014-0988-2},
	language = {en},
	number = {1},
	urldate = {2026-08-21},
	journal = {Journal of Engineering Physics and Thermophysics},
	author = {Dalir, N. and Nourazar, S. S.},
	month = jan,
	year = {2014},
	pages = {89--97},
}

@article{abu-hijleh_analytical_1997,
	title = {Analytical {Solution} of {Heat} {Conduction} in a {Multi}-{Layer} {Orthotropic} {Cylinder} {Subject} to {Periodic} and {Asymmetric} {Heat} {Flux}},
	volume = {9},
	copyright = {https://www.elsevier.com/tdm/userlicense/1.0/},
	issn = {10183639},
	url = {https://linkinghub.elsevier.com/retrieve/pii/S1018363918306809},
	doi = {10.1016/S1018-3639(18)30680-9},
	language = {en},
	number = {2},
	urldate = {2026-08-21},
	journal = {Journal of King Saud University - Engineering Sciences},
	author = {Abu-Hijleh, Bassam A.},
	year = {1997},
	pages = {251--263},
}

@article{ozisik_transient_1980,
	title = {Transient heat conduction in an anisotropic medium in cylindrical coordinates},
	volume = {309},
	issn = {00160032},
	url = {https://linkinghub.elsevier.com/retrieve/pii/0016003280900952},
	doi = {10.1016/0016-0032(80)90095-2},
	language = {en},
	number = {6},
	urldate = {2026-08-21},
	journal = {Journal of the Franklin Institute},
	author = {Özişik, M.N. and Shouman, S.M.},
	month = jun,
	year = {1980},
	pages = {457--472},
}

@article{jiang_new_2022,
	title = {A new spatial-domain thermoreflectance method to measure a broad range of anisotropic in-plane thermal conductivity},
	volume = {191},
	issn = {00179310},
	url = {https://linkinghub.elsevier.com/retrieve/pii/S0017931022003313},
	doi = {10.1016/j.ijheatmasstransfer.2022.122849},
	language = {en},
	urldate = {2024-08-12},
	journal = {International Journal of Heat and Mass Transfer},
	author = {Jiang, Puqing and Wang, Dihui and Xiang, Zeyu and Yang, Ronggui and Ban, Heng},
	month = aug,
	year = {2022},
	pages = {122849},
}

@article{wang_spatially_2024,
	title = {Spatially resolved lock-in micro-thermography ({SR}-{LIT}): {A} tensor analysis-enhanced method for anisotropic thermal characterization},
	volume = {11},
	issn = {1931-9401},
	shorttitle = {Spatially resolved lock-in micro-thermography ({SR}-{LIT})},
	url = {https://pubs.aip.org/apr/article/11/2/021407/3282197/Spatially-resolved-lock-in-micro-thermography-SR},
	doi = {10.1063/5.0191073},
	language = {en},
	number = {2},
	urldate = {2024-06-28},
	journal = {Applied Physics Reviews},
	author = {Wang, Dihui and Ban, Heng and Jiang, Puqing},
	month = jun,
	year = {2024},
	pages = {021407},
}

@article{wang_three-dimensional_2025,
	title = {Three-dimensional ({3D}) tensor-based methodology for characterizing {3D} anisotropic thermal conductivity tensor},
	volume = {242},
	issn = {00179310},
	url = {https://linkinghub.elsevier.com/retrieve/pii/S0017931025002273},
	doi = {10.1016/j.ijheatmasstransfer.2025.126886},
	language = {en},
	urldate = {2025-04-28},
	journal = {International Journal of Heat and Mass Transfer},
	author = {Wang, Dihui and Ban, Heng and Jiang, Puqing},
	month = jun,
	year = {2025},
	pages = {126886},
}

@article{sheikh_photothermal_2025,
	title = {Photothermal radiometry dual measurement technique to reduce uncertainty and improve convergence},
	volume = {169},
	issn = {07351933},
	url = {https://linkinghub.elsevier.com/retrieve/pii/S0735193325013089},
	doi = {10.1016/j.icheatmasstransfer.2025.109882},
	language = {en},
	urldate = {2026-05-18},
	journal = {International Communications in Heat and Mass Transfer},
	author = {Sheikh, Raza and Pompidou, Quentin and Wang, Dihui and Horny, Nicolas and Ban, Heng},
	month = dec,
	year = {2025},
	pages = {109882},
}

@article{wang_photothermal_2004,
	title = {Photothermal {Radiometry} with {Solid} {Cylindrical} {Samples}},
	language = {en},
	author = {Wang, Chinhua and Mandelis, Andreas and Liu, Yue},
	year = {2004},
}

@article{wang_thermal-wave_2005,
	title = {Thermal-wave nondestructive evaluation of cylindrical composite structures using frequency-domain photothermal radiometry},
	volume = {97},
	issn = {0021-8979, 1089-7550},
	url = {https://pubs.aip.org/jap/article/97/1/014911/897389/Thermal-wave-nondestructive-evaluation-of},
	doi = {10.1063/1.1819999},
	language = {en},
	number = {1},
	urldate = {2025-12-17},
	journal = {Journal of Applied Physics},
	author = {Wang, Chinhua and Mandelis, Andreas and Liu, Yue},
	month = jan,
	year = {2005},
	pages = {014911},
}

@article{salazar_propagation_2006,
	title = {Propagation of thermal waves in multilayered cylinders using the thermal quadrupole method},
	language = {en},
	author = {Salazar, A and Celorrio, R},
    year = {2006},
}

@article{xie_thermal_2011,
	title = {Thermal conductivity depth-profile reconstruction of multilayered cylindrical solids using the thermal-wave {Green} function method},
	language = {en},
	author = {Xie, Guangxi and Zhang, Jie and Liu, Liwang and Wang, Chinhua and Mandelis, Andreas},
	year = {2011},
}

@article{liu_quantitative_2020,
	title = {Quantitative photothermal lock-in thermography imaging of curved surfaces of cylindrical solids},
	volume = {127},
	issn = {0021-8979, 1089-7550},
	url = {https://pubs.aip.org/jap/article/127/19/195101/1065115/Quantitative-photothermal-lock-in-thermography},
	doi = {10.1063/5.0007779},
	language = {en},
	number = {19},
	urldate = {2025-09-18},
	journal = {Journal of Applied Physics},
	author = {Liu, Yu and Sun, Ti and Wang, Chinhua and Mandelis, Andreas},
	month = may,
	year = {2020},
	pages = {195101},
}

@article{feser_pump-probe_2014,
	title = {Pump-probe measurements of the thermal conductivity tensor for materials lacking in-plane symmetry},
	volume = {85},
	issn = {0034-6748, 1089-7623},
	url = {https://pubs.aip.org/rsi/article/85/10/104903/361912/Pump-probe-measurements-of-the-thermal},
	doi = {10.1063/1.4897622},
	language = {en},
	number = {10},
	urldate = {2024-10-03},
	journal = {Review of Scientific Instruments},
	author = {Feser, Joseph P. and Liu, Jun and Cahill, David G.},
	month = oct,
	year = {2014},
	pages = {104903},
}

@article{tang_anisotropic_2021,
	title = {Anisotropic thermal conductivity tensor measurements using beam-offset frequency domain thermoreflectance ({BO}-{FDTR}) for materials lacking in-plane symmetry},
	volume = {164},
	issn = {00179310},
	url = {https://linkinghub.elsevier.com/retrieve/pii/S0017931020335365},
	doi = {10.1016/j.ijheatmasstransfer.2020.120600},
	language = {en},
	urldate = {2024-07-21},
	journal = {International Journal of Heat and Mass Transfer},
	author = {Tang, Lei and Dames, Chris},
	month = jan,
	year = {2021},
	pages = {120600},
}

@article{rodin_simultaneous_2017,
	title = {Simultaneous measurement of in-plane and through-plane thermal conductivity using beam-offset frequency domain thermoreflectance},
	volume = {88},
	issn = {0034-6748, 1089-7623},
	url = {https://pubs.aip.org/rsi/article/88/1/014902/367479/Simultaneous-measurement-of-in-plane-and-through},
	doi = {10.1063/1.4973297},
	language = {en},
	number = {1},
	urldate = {2024-07-21},
	journal = {Review of Scientific Instruments},
	author = {Rodin, David and Yee, Shannon K.},
	month = jan,
	year = {2017},
	pages = {014902},
}

@article{deck_demonstration_nodate,
	title = {{DEMONSTRATION} {OF} {ENGINEERED} {MULTI}-{LAYERED} {SIC}- {SIC} {CLADDING} {WITH} {ENHANCED} {ACCIDENT} {TOLERANCE}},
	language = {en},
	author = {Deck, C P and Khalifa, H E and Jacobsen, G and Sheeder, J and Shapovalov, K and Song, E and Gazza, J and Back, C A and Xu, P and Boylan, F and Jacko, R},
}

@article{deck_characterization_2015,
	title = {Characterization of {SiC}–{SiC} composites for accident tolerant fuel cladding},
	volume = {466},
	issn = {00223115},
	url = {https://linkinghub.elsevier.com/retrieve/pii/S0022311515301598},
	doi = {10.1016/j.jnucmat.2015.08.020},
	language = {en},
	urldate = {2025-12-12},
	journal = {Journal of Nuclear Materials},
	author = {Deck, C.P. and Jacobsen, G.M. and Sheeder, J. and Gutierrez, O. and Zhang, J. and Stone, J. and Khalifa, H.E. and Back, C.A.},
	month = nov,
	year = {2015},
	pages = {667--681},
}

@article{wang_anisotropic_2022,
	title = {Anisotropic thermal diffusivity and conductivity in {SiC}/{SiC} tubes studied by infrared imaging and {X}-ray computed tomography},
	volume = {48},
	issn = {02728842},
	url = {https://linkinghub.elsevier.com/retrieve/pii/S027288422201327X},
	doi = {10.1016/j.ceramint.2022.04.153},
	language = {en},
	number = {15},
	urldate = {2026-08-28},
	journal = {Ceramics International},
	author = {Wang, Hsin and Koyanagi, Takaaki and Arregui-Mena, D.J. and Katoh, Yutai},
	month = aug,
	year = {2022},
	pages = {21717--21727},
}

@article{fave_radial_2017,
	title = {A radial heat flow apparatus for thermal conductivity characterisation of cylindrical samples},
	volume = {130},
	issn = {1388-6150, 1588-2926},
	url = {http://link.springer.com/10.1007/s10973-017-6578-8},
	doi = {10.1007/s10973-017-6578-8},
	language = {en},
	number = {3},
	urldate = {2024-06-10},
	journal = {Journal of Thermal Analysis and Calorimetry},
	author = {Fave, Loïc and Pouchon, Manuel A. and Hébert, Cécile},
	month = dec,
	year = {2017},
	pages = {1855--1863},
}

@article{he_method_2023,
	title = {Method to {Measure} {Radial} {Thermal} {Conductivity} for {Cylindrical} {Samples}},
	volume = {8},
	copyright = {https://creativecommons.org/licenses/by-nc-nd/4.0/},
	issn = {2470-1343, 2470-1343},
	url = {https://pubs.acs.org/doi/10.1021/acsomega.2c06901},
	doi = {10.1021/acsomega.2c06901},
	language = {en},
	number = {7},
	urldate = {2026-08-28},
	journal = {ACS Omega},
	author = {He, Jiahuan and Wen, Long and He, Xiao and He, Tingting and Liu, Tangyan and Kang, Qiang and Wang, Daocheng and Peng, Xian and Yao, Hongyu and Zhang, Yin and Deng, Xiaohang},
	month = feb,
	year = {2023},
	pages = {6530--6537},
}

@article{corona_recent_2023,
	title = {Recent {Progress} in {Modulated} {Photothermal} {Radiometry}},
	volume = {23},
	issn = {1424-8220},
	url = {https://www.mdpi.com/1424-8220/23/10/4935},
	doi = {10.3390/s23104935},
	language = {en},
	number = {10},
	urldate = {2025-11-13},
	journal = {Sensors},
	author = {Corona, Javier and Kandadai, Nirmala},
	month = may,
	year = {2023},
	pages = {4935},
}

@article{salazar_thermal_2024,
	title = {Thermal diffusivity of rods, tubes, and spheres by the ﬂash method},
	language = {en},
	author = {Salazar, Agustín and Garrido, Florencio and Celorrio, Ricardo},
	year = {2024},
}

@article{duquesne_flash_2016,
	title = {A flash characterisation method for thin cylindrical multilayered composites based on the combined front and rear faces thermograms},
	volume = {13},
	issn = {1768-6733, 2116-7176},
	url = {https://www.tandfonline.com/doi/full/10.1080/17686733.2016.1165958},
	doi = {10.1080/17686733.2016.1165958},
	language = {en},
	number = {2},
	urldate = {2025-05-01},
	journal = {Quantitative InfraRed Thermography Journal},
	author = {Duquesne, Loys and Lorrette, Christophe and Pradère, Christophe and Vignoles, Gérard Louis and Batsale, Jean-Christophe},
	month = jul,
	year = {2016},
	pages = {182--194},
}

@article{salazar_thermal_2011,
	title = {Thermal diffusivity of nonflat plates using the flash method},
	volume = {82},
	issn = {0034-6748, 1089-7623},
	url = {https://pubs.aip.org/rsi/article/82/1/014902/910980/Thermal-diffusivity-of-nonflat-plates-using-the},
	doi = {10.1063/1.3529431},
	language = {en},
	number = {1},
	urldate = {2026-08-28},
	journal = {Review of Scientific Instruments},
	author = {Salazar, Agustín and Fuente, Raquel and Apiñaniz, Estibaliz and Mendioroz, Arantza},
	month = jan,
	year = {2011},
	pages = {014902},
}

@article{jiang_tutorial_2018,
	title = {Tutorial: {Time}-domain thermoreflectance ({TDTR}) for thermal property characterization of bulk and thin film materials},
	volume = {124},
	issn = {0021-8979, 1089-7550},
	shorttitle = {Tutorial},
	url = {https://pubs.aip.org/jap/article/124/16/161103/1030095/Tutorial-Time-domain-thermoreflectance-TDTR-for},
	doi = {10.1063/1.5046944},
	language = {en},
	number = {16},
	urldate = {2024-09-23},
	journal = {Journal of Applied Physics},
	author = {Jiang, Puqing and Qian, Xin and Yang, Ronggui},
	month = oct,
	year = {2018},
	pages = {161103},
}

@article{popov_characterization_1999,
	title = {Characterization of rock thermal conductivity by high-resolution optical scanning},
	volume = {28},
	issn = {03756505},
	url = {https://linkinghub.elsevier.com/retrieve/pii/S0375650599000073},
	doi = {10.1016/S0375-6505(99)00007-3},
	language = {en},
	number = {2},
	urldate = {2026-08-28},
	journal = {Geothermics},
	author = {Popov, Yuri A. and Pribnow, Dan F.C. and Sass, John H. and Williams, Colin F. and Burkhardt, Hans},
	month = apr,
	year = {1999},
	pages = {253--276},
}

@article{vardi_thermal_1963,
	title = {Thermal {Conductivity} of {Anisotropic} {Solids} at {High} {Temperatures}—{The} {Thermal} {Conductivity} of {Molded} and {Pyrolytic} {Graphites}},
	volume = {46},
	copyright = {http://onlinelibrary.wiley.com/termsAndConditions\#vor},
	issn = {0002-7820, 1551-2916},
	url = {https://ceramics.onlinelibrary.wiley.com/doi/10.1111/j.1151-2916.1963.tb11695.x},
	doi = {10.1111/j.1151-2916.1963.tb11695.x},
	language = {en},
	number = {3},
	urldate = {2026-08-28},
	journal = {Journal of the American Ceramic Society},
	author = {Vardi, Joseph and Hoch, Michael},
	month = mar,
	year = {1963},
	pages = {129--132},
}

@article{vay_anisotropic_2026,
	title = {Anisotropic thermal conductivity as a function of wood density: experimental data comparison across eight wood species},
	volume = {80},
	copyright = {http://creativecommons.org/licenses/by/4.0},
	issn = {0018-3830, 1437-434X},
	shorttitle = {Anisotropic thermal conductivity as a function of wood density},
	url = {https://www.degruyterbrill.com/document/doi/10.1515/hf-2025-0097/html},
	doi = {10.1515/hf-2025-0097},
	language = {en},
	number = {2},
	urldate = {2026-08-28},
	journal = {Holzforschung},
	author = {Vay, Oliver and Hansmann, Christian},
	month = feb,
	year = {2026},
	pages = {115--124},
}

\end{document}